\documentclass[aps,pra,reprint,superscriptaddress,longbibliography,noeprint]{revtex4-1} 
\usepackage[colorlinks,pdfstartview=FitH,citecolor=blue,linkcolor=blue,urlcolor=blue]{hyperref} 
\usepackage{graphicx}
\usepackage{amsmath}
\usepackage{amssymb}
\usepackage{siunitx}
\usepackage{bm} 
\usepackage{mhchem} 

\renewcommand{\Re}{\operatorname{Re}}

\newcommand{\citeasnoun}[1]{Ref.~\cite{#1}} 

\renewcommand{\eqref}[1]{Eq.~(\ref{eq:#1})}
\newcommand{\eqreftwo}[2]{Eqs.~(\ref{eq:#1},\ref{eq:#2})}
\newcommand{\eqrefrange}[2]{Eqs.~(\ref{eq:#1}--\ref{eq:#2})}

\newcommand{\Eqreftwo}[2]{Equations~(\ref{eq:#1},\ref{eq:#2})}
\newcommand{\figref}[1]{Fig.~\ref{fig:#1}}
\newcommand{\Figref}[1]{Figure~\ref{fig:#1}}
\newcommand{\cc}[1]{#1^*}
\newcommand{\vect}[1]{\mathbf{#1}}

\newcommand*{\Pabs}{P_{\rm abs}}

\newcommand*{\Ev}{\vect{E}}

\newcommand*{\Hv}{\vect{H}}
\newcommand*{\xv}{\vect{x}}

\newcommand*{\cv}{\vect{c}}
\newcommand*{\cin}{\vect{c}_{\rm in}}
\newcommand*{\cout}{\vect{c}_{\rm out}}
\newcommand*{\Vbb}{\mathbb{V}}
\newcommand*{\Vin}{\Vbb_{-}}
\newcommand*{\Vout}{\Vbb_{+}}
\newcommand*{\Sbb}{\mathbb{S}}
\newcommand*{\Qbb}{\mathbb{Q}}
\newcommand*{\Pbb}{\mathbb{P}}
\newcommand*{\Jbb}{\mathbb{J}}

\newcommand*{\Wc}{\mathcal{W}}
\newcommand*{\Pc}{\mathcal{P}}
\newcommand*{\Jc}{\mathcal{J}}
\newcommand*{\Win}{\Wc_{\rm in}}
\newcommand*{\Pin}{\Pc_{\textrm{in},i}}
\newcommand*{\Jin}{\Jc_{\textrm{in},i}}
\newcommand*{\Mlm}{\vect{M}^{+/-}_{\ell,m}}
\newcommand*{\Nlm}{\vect{N}^{+/-}_{\ell,m}}
\newcommand*{\lmax}{\ell_{\rm max}}
\newcommand*{\psiin}{\psi_{\rm in}}
\newcommand*{\psiout}{\psi_{\rm out}}

\newcommand*{\SM}{SI}
\newcommand*{\nhat}{\hat{\vect{n}}}

\begin{document} 

\title{Optimal nanoparticle forces, torques, and illumination fields}

\author{Yuxiang Liu}
\affiliation{Department of Applied Physics and Energy Sciences Institute, Yale University, New Haven, CT 06511}
\author{Lingling Fan}
\affiliation{Department of Applied Physics and Energy Sciences Institute, Yale University, New Haven, CT 06511}
\affiliation{School of Physics and National Laboratory of Solid State Microstructures, Nanjing University, Nanjing 210093, China}
\author{Yoonkyung E. Lee}
\affiliation{Department of Mechanical Engineering, Massachusetts Institute of Technology, Cambridge, MA 02139}
\author{Nicholas X. Fang}
\affiliation{Department of Mechanical Engineering, Massachusetts Institute of Technology, Cambridge, MA 02139}
\author{Steven G. Johnson}
\affiliation{Department of Physics, Massachusetts Institute of Technology, Cambridge, MA 02139}
\affiliation{Department of Mathematics, Massachusetts Institute of Technology, Cambridge, MA 02139}
\author{Owen D. Miller}
\email[Corresponding author: ]{owen.miller@yale.edu}  
\affiliation{Department of Applied Physics and Energy Sciences Institute, Yale University, New Haven, CT 06511}


\begin{abstract}
A universal property of resonant subwavelength scatterers is that their optical cross-sections are proportional to a square wavelength, $\lambda^2$, regardless of whether they are plasmonic nanoparticles, two-level quantum systems, or RF antennas. The maximum cross-section is an intrinsic property of the \emph{incident field}: plane waves, with infinite power, can be decomposed into multipolar orders with finite powers proportional to $\lambda^2$. In this Article, we identify $\lambda^2/c$ and $\lambda^3/c$ as analogous force and torque constants, derived within a more general quadratic scattering-channel framework for upper bounds to optical force and torque for any illumination field. This framework also solves the reverse problem: computing globally optimal ``holographic'' incident beams, for a fixed collection of scatterers. We analyze structures and incident fields that approach the bounds, which for wavelength-scale bodies show a rich interplay between scattering channels, and we show that spherically symmetric structures are forbidden from reaching the plane-wave force/torque bounds. This framework should enable optimal mechanical control of nanoparticles with light.
\end{abstract}

\maketitle
\sisetup{range-phrase=--}
\sisetup{range-units=single}

Optically induced forces and torques offer precise mechanical control of nanoparticles~\cite{Grier2003,Neuman2004,Hanggi2009,Juan2011,Ilic2016}, yet a basic understanding of what is possible has been limited by the inherent complexity in the optical response of a nanoparticle of any size, shape, and material. Here we show that a general scattering-channel decomposition embeds optical-response functions into matrix quadratic forms that, in tandem with a convex passivity constraint, readily yield \emph{analytical} upper bounds for scatterers under arbitrary illumination. For plane waves, the force and torque bounds are proportional to $\lambda^2/c$ and $\lambda^3/c$, respectively (for wavelength $\lambda$ and speed of light $c$), for scatterers of any size, analogous to the well-known $\sim\lambda^2$ cross-section of a small scatterer~\cite{Hamam2007,Ruan2011,Loudon2000,Stutzman2012,Pozar2009,Liberal2014}. Spheres, cylinders, and helices can approach the various bounds, which often require a complex interplay between scattering channels. With modern progress in spatial light modulators~\cite{Arrizon2007,Hernandez-Hernandez2010,Clark2016} and other beam-shaping techniques~\cite{Karimi2009,Schulz2013,Chen2015}, the ``reverse'' problem of shaping the incident field for a fixed geometry is increasingly important. Our quadratic-form framework naturally yields \emph{globally} optimal illumination fields as extremal eigenvectors of Hermitian matrices. For a generic scattering problem, we show that optimized incident fields can achieve sizable enhancements (20--40$\times$) to optically induced force and torque, offering orders-of-magnitude enhancements over conventional beams.

Mechanical forces induced by light are the foundation for optical trapping and manipulation---versatile tools with applications ranging from laser cooling~\cite{Metcalf1999} and nanoparticle guidance~\cite{Ashkin1970,Allen2003,Bonin2002,Pelton2006,Tong2010,Dholakia2011,Wang2011d,Ilic2016} to biomolecular sensing~\cite{Eriksson2007,Jonas2008,Wang2011c}. In the limit of dipolar response, analytical expressions for force and torque are known, as are associated concepts such as ``gradient'' forces~\cite{Ashkin1986,Nieto-Vesperinas2010,Jones2015} and optical ``chirality''~\cite{Barron2004,Tang2010,Bliokh2011}. At wavelength size scales and larger, the only structures for which analytical bounds or semianalytical response expressions are known are ray-optical~\cite{Ashkin1992,Jones2015} or spherical~\cite{Rahimzadegan2017}. For nonspherical scatterers, optical forces and torques generally require simulation of Maxwell's equations~\cite{Waterman1971,Mishchenko1996,Doicu2006,Nieminen2007,Reid2015}, providing numerical results but little insight. This contrasts strongly with the more detailed knowledge of \emph{power} flow in such systems, ranging from bounds~\cite{Hamam2007,Ruan2011,Pozar2009,Liberal2014,Miller2016,Yang2017,Miller2017} to sum rules~\cite{Gordon1963,Purcell1969,Sohl2007a} to spherical-particle design criteria~\cite{Grigoriev2015}. The disparity between the broad understanding of power flow versus the relative paucity for momentum flow may reflect the complexity of the Maxwell stress tensor relative to the Poynting vector. But as we show below, for passive systems in which energy is not supplied to the polarization currents, the requirement that outgoing power is less than incoming power is a convex constraint dictating what is possible for power, momentum, and other quantities of interest.

``Holographic'' optical force and torque generation~\cite{Polin2005,Grier2006,Martin-Badosa2007,Bianchi2010,Taylor2015} faces similar challenges. Whereas analytical bounds can be derived for the concentration of light for power transfer~\cite{Sheppard1994}, especially for dipolar objects~\cite{Gerhardt2007,Zumofen2008}, finding optimal illumination fields for force/torque typically requires iterative computational optimization schemes~\cite{Polin2005,Lee2017,Taylor2017} which may not converge to a global optimum. Recent work has identified the potential of quadratic forms for phase optimization~\cite{Taylor2017}, ``absorption''-like energy-exchange quantities~\cite{Fernandez-Corbaton2017}, or ``optical eigenmodes''~\cite{Mazilu2011}; the framework here shows generally how quadratic frameworks enable global optimization for any power quantity.

\citeasnoun{Fernandez-Corbaton2017} recently developed a framework that complements the one we use below. The authors identify conservation laws for ``transfer'' quantities---such as absorbed power, force, and torque---and derive upper bounds for such transfer rates. The bounds they derive for a given object are quite different from the analytical bounds which we derive for arbitrary scatterers: their bounds require the full scattering matrix of an object (as no other constraints are considered), whereas we use passivity as a constraint and derive bounds without knowledge of the scattering matrix, requiring only the number of incoming channels for which there is non-trivial coupling. The illumination bounds of \citeasnoun{Fernandez-Corbaton2017} are closer to the bounds we derive for optimal illumination, with the key difference that our bounds apply generally to scattering quantities (scattered/extinguished power, linear momentum, angular momentum, etc.) that may not be ``transfer'' properties but which are necessarily quadratic forms.

\section*{Scattering-Channel Framework}
The scattering properties of a body are uniquely determined by the incoming and outgoing fields on any bounding surface~\cite{Jin2011}. We represent all electromagnetic fields in six-dimensional tensors,
\begin{align}
    \psi = \begin{pmatrix}
        \Ev \\
        \Hv
    \end{pmatrix}.
\end{align}
For a fixed frequency $\omega$ (time-dependence $e^{-i\omega t}$), the ``scattering channels'' are basis sets on or outside a bounding surface of all scatterers in a given problem; equivalently, they are the ``ports'' commonly used in temporal coupled-mode theory~\cite{Newton2013,Suh2004}. We assume that the surface encloses all scatterers (such that all channels are propagating or far-field in nature), that the background is lossless, so that each channel carries fixed and position-independent energy and momenta, and that a finite set of channels describe the scattering process with arbitrarily high accuracy. We start by considering a basis of $N$ ``incoming'' channels, represented by basis states $\varphi_{1-}$ through $\varphi_{N-}$ in a tensor $\Vin$:
\begin{align}
    \Vin = \begin{pmatrix}
        \varphi_{1-}(\xv) & \varphi_{2-}(\xv) & \ldots & \varphi_{N-}(\xv)
    \end{pmatrix}.
\end{align}
Any complete set of incident channels may be used (plane waves, vector spherical waves, etc.). For the analytical force/torque bounds we will derive, the incident field will be fixed for a given problem, whereas for the illumination-field bounds, it will comprise the degrees of freedom to be optimized, in which case it is always possible to constrain the illumination to a subset of solid angles, as may be experimentally advantageous. From here, we will show how to construct sets of power-orthogonal incoming and outgoing states, and that for any energy/momentum quantity there is a certain orthogonality between incoming and outgoing states that simplifies the ultimate quadratic forms.

The power flowing \emph{into} a surface $S$ with outward normal $\nhat$ is given by
\begin{align}
    -\frac{1}{2} \Re \int_S \Ev \times \cc{\Hv} \cdot \nhat = -\frac{1}{4} \int_S \psi^\dagger \underbrace{\begin{pmatrix} & -\nhat \times \\ \nhat \times & \end{pmatrix}}_\Theta \psi,
\end{align}
where $\Theta$ is a real-symmetric matrix (cross products change sign under interchange of their arguments, so $\left(\nhat \times\right)^T = -\nhat \times$ and vice versa). For the incoming-wave basis $\Vin$, with linearly independent but not necessarily orthonormal states, the power of an incoming field $\psi_{\rm in}$ in this basis, $\psi_{\rm in} = \Vin \vect{c}_{\rm in}$, is
\begin{align}
    \cv_{\rm in} \left[ \int_S \Vin^\dagger \left(-\frac{1}{4} \Theta\right) \Vin \right] \cv_{\rm in}.
\end{align}
Now we use the physical knowledge that $\Vin$ comprises only \emph{incoming} states (with nonzero power) to assert that $-\Theta/4$ is \emph{positive-definite} over all states of interest. Since $-\Theta/4$ is definite, it can be used to define a modified inner product, and then one can use e.g. the Gram-Schmidt process to orthonormalize our $\Vin$ basis in this quadratic form~\cite{Horn2013}, giving:
\begin{align}
    \int_S \Vin^\dagger \left( -\frac{1}{4} \Theta \right) \Vin = \mathcal{I}.
    \label{eq:ortho}
\end{align}
(Note that if the ambient medium is periodic, the surface $S$ needs to be replaced by a volume that is one unit cell thick~\cite{Johnson2002a}. The Bloch waves in a periodic medium will not be linearly independent on a single cross-section.)

For power-orthogonal outgoing channels, we \emph{time-reverse} the incoming channels. The outgoing channels, denoted $\Vout$, are then given by
\begin{align}
    \Vout = \underbrace{\begin{pmatrix}
        \mathcal{I} & \\
         & -\mathcal{I}
    \end{pmatrix}}_\mathcal{P} \cc{\Vin},
    \label{eq:symm}
\end{align}
where the parity matrix $\mathcal{P}$ accounts for the different time-reversal properties of electric ($\Ev~\rightarrow~\cc{\Ev}$) and magnetic ($\Hv~\rightarrow~-\cc{\Hv}$) fields. These states have the opposite normalization, because the power is flowing in the opposite direction: $\int_S \Vout^\dagger \left(-\frac{1}{4} \Theta\right) \Vout = \int_S \Vin^T \mathcal{P} \left(-\frac{1}{4} \Theta\right) \mathcal{P} \cc{\Vin} = -\int_S \Vin^T \left(-\frac{1}{4} \Theta\right) \cc{\Vin} = - \cc{\int_S \Vin^\dagger \left(-\frac{1}{4} \Theta\right) \Vin} = -\mathcal{I}$, where we used the fact that $\mathcal{P} \Theta \mathcal{P} = -\Theta$, as can be verified by direct substitution. Thus we have constructed power-orthonormal sets of incoming and outgoing states.

Any non-trivial field solution of a scattering solution will comprise both incoming and outgoing waves, and thus computing power, force, torque, or another quadratic form will include ``overlap'' terms between the incoming states of $\Vin$ and the outgoing states of $\Vout$. We can show generally that such terms will always cancel. Consider an energy/momentum-flux quantity that is a quadratic form of the fields flowing through a surface $S$:
\begin{align}
    Q = \int_S \psi^\dagger \mathcal{Q} \psi,
    \label{eq:quadform}
\end{align}
where $\mathcal{Q}$ is a Hermitian operator determined by, e.g., the Poyting vector or the electromagnetic stress tensor. In the SM, we use the time-reversal pairing of the input/output states to prove that $\mathcal{Q}$ must satisfy the time-reversal expression
\begin{align}
    \mathcal{Q} = -\mathcal{P} \mathcal{Q}^T \mathcal{P}.
    \label{eq:Q_TR}
\end{align}
To show orthogonality between the incoming and outgoing waves, we now consider a scenario in which no absorption occurs, and all incoming power/momentum is converted into outgoing power/momentum. The total fields are given by the incoming and outgoing fields
\begin{align}
    \psi &= \psi_{\rm in} + \psi_{\rm out} \nonumber \\
    &= \Vin \cin + \Vout \cout.
\end{align}
Evaluating the power/momentum quantity on the surface $S$, we find
\begin{align}
    Q &= \int_S \left[ \psiin^\dagger \mathcal{Q} \psiin + \psiout^\dagger \mathcal{Q} \psiout + 2 \Re \psiin^\dagger \mathcal{Q} \psiout \right] \nonumber \\
    &= \cin^\dagger \left( \int_S \Vin^\dagger \mathcal{Q} \Vin \right) \cin + \cout^\dagger \left( \int_S \Vout^\dagger \mathcal{Q} \Vout \right) \cout  \nonumber \\
    &+ 2\Re\left[\cin^\dagger \left( \int_S \Vin^\dagger \mathcal{Q} \Vout \right) \cout \right].
    \label{eq:Qall}
\end{align}
From \eqref{Q_TR}, it is straightforward to show that incoming/outgoing channels carry equal and opposite energy/momentum: $\int_S \Vout^\dagger \mathcal{Q} \Vout = - \int_S \Vin^\dagger \mathcal{Q} \Vin$. Thus the first two terms in \eqref{Qall} add to zero, and we are left only with the third term. The total sum $Q$ has to equal zero, leaving
\begin{align}
    \Re \int_S \Vin^\dagger \mathcal{Q} \Vout = 0.
    \label{eq:InOutOverlap}
\end{align}
Thus the time-reversed, propagating basis states exhibit orthogonality between incoming and outgoing waves, for any flux quantity represented by a quadratic form.

\section*{Analytical bounds}
By virtue of linearity, any quantity describing energy or momentum flow of a field $\psi$ through a surface $S$ can be described by \eqref{quadform}, as a quadratic form.\cite{Jackson1999,Reid2015} We can decompose the incoming- and outgoing-wave components of $\psi$ into basis-coefficient vectors $\cin$ and $\cout$:
\begin{subequations}
\begin{align}
    \psi_{\rm in}(\xv) &= \Vin(\xv) \cin \\
    \psi_{\rm out}(\xv) &= \Vout(\xv) \cout.
\end{align}
\end{subequations}
For linear materials (considered hereafter), the basis coefficients $\cin$ and $\cout$ are related by $\cout = \Sbb \cin$, where $\Sbb$ is the scattering matrix. By the power-orthonormalization condition for $\Vin$, \eqref{ortho}, and its negative for $\Vout$, absorption is simply
\begin{align}
    \Pabs = \cin^\dagger \cin - \cout^\dagger \cout,
    \label{eq:power}
\end{align}
    i.e., incoming minus outgoing power. Similarly, the force or torque on any scatterer, in some direction $i$, is the difference in momentum flux of the incoming and outgoing waves, given by
\begin{align}
    F_i &= \frac{1}{c} \left[\cin^\dagger \Pbb_i \cin - \cout^\dagger \Pbb_i \cout\right], \label{eq:force} \\
    \tau_i &= \frac{1}{\omega} \left[\cin^\dagger \Jbb_i \cin - \cout^\dagger \Jbb_i \cout\right], \label{eq:torque}
\end{align}
where $c$ is the speed of light, and $\Pbb_i$ and $\Jbb_i$ are dimensionless matrix measures of linear and angular momentum, given by overlap integrals (described above) involving the stress tensor (\SM). There are no cross terms, as proven by \eqref{InOutOverlap}. \eqrefrange{power}{torque} compactly represent enegy/momentum flow in an intuitive basis. We can derive general bounds by adding a single constraint: passivity.

\begin{figure}
    \centering
    \includegraphics[width=\linewidth]{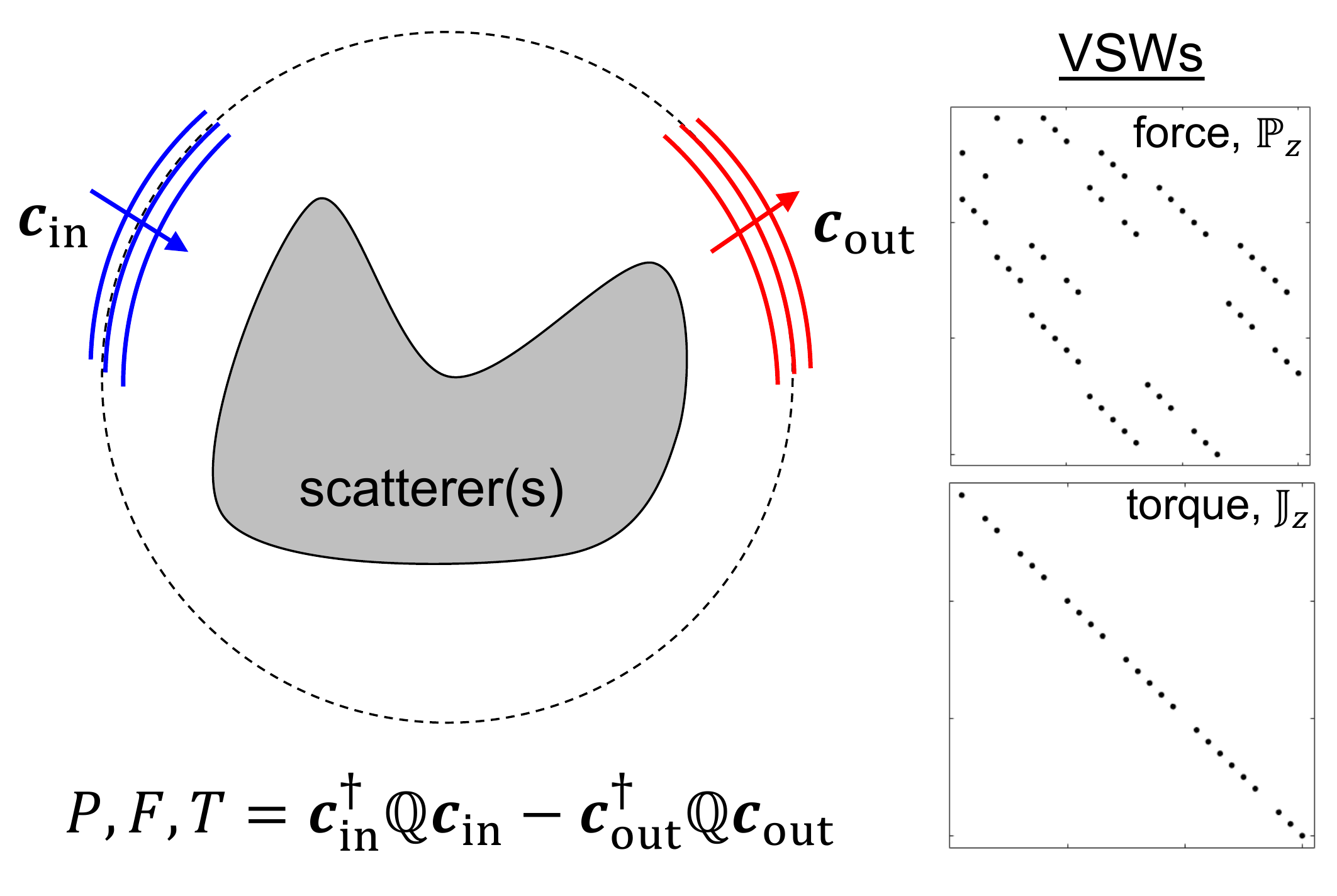}
    \caption{The power, force, or torque imparted to any structure (or collection thereof) can be encoded in matrix quadratic forms $\Qbb$ that are amenable to analytical bounds and quadratic optimization. In a vector-spherical-wave (VSW) basis, the force ($\Qbb \rightarrow \Pbb_z$) and torque ($\Qbb \rightarrow \Jbb_z$) matrices have nonzero values as shown on the right.} 
    \label{fig:fig1}
\end{figure}

Passivity requires that induced currents do no work~\cite{Welters2014}; as a consequence, absorption and scattered power are nonnegative. In a recent series of papers~\cite{Miller2015,Miller2016,Yang2017,Miller2017,Yang2018,Shim2018subm}, we have identified passivity-based quadratic constraints to the currents induced within a medium, and applied them to find material-dictated bounds to a variety of optical-response functions. Here, we apply such constraints to the scattering channels themselves. Nonnegative absorption, i.e. $\Pabs > 0$, translates \eqref{power} to a quadratic photon-conservation constraint on $\cout$:
\begin{align}
    \cout^\dagger \cout \leq \cin^\dagger \cin.
    \label{eq:passive}
\end{align}

The largest force or torque that can be exerted on a nanoparticle can thus be formulated as the maximum of \eqreftwo{force}{torque} subject to passivity, i.e. \eqref{passive}. \eqrefrange{force}{passive} represent a particularly straightforward quadratic optimization with quadratic constraints. Of the two terms each in \eqreftwo{force}{torque}, the first are fixed by the incident field, while the second are the variable ones to be bounded.  For simplicity, we assume the standard case in which channels have equal positive- and negative-momentum eigenstates, such that the eigenvalues come in positive/negative pairs and $\max [\cout^\dagger (-\Qbb) \cout] = \max [\cout^\dagger \Qbb \cout]$ (it is straightforward to generalize the results for alternative bases). Then the Rayleigh quotient~\cite{Horn2013} in tandem with the passivity constraint, \eqref{passive}, bounds the second terms of \eqreftwo{force}{torque} by $\cout^\dagger \Qbb \cout \leq (\cout^\dagger \cout) \lambda_{\rm max}(\Qbb) \leq (\cin^\dagger \cin) \lambda_{\rm max}(\Qbb)$, for $\Qbb = \Pbb_i,\Jbb_i$, where $\lambda_{\rm max}(\Qbb)$ is the largest eigenvalue of $\Qbb$. Denoting the incoming power, momentum flow, and angular momentum flow by $\Win = \cin^\dagger \cin$, $\Pin = \cin^\dagger \Pbb_i \cin / c$, and $\Jin = \cin^\dagger \Jbb_i \cin / \omega$, respectively, the maximum force and torque are given by
\begin{align}
    F_i &\leq \Pin + \frac{\Win}{c} \lambda_{\rm max}(\Pbb_i) \label{eq:force_bound}, \\
    \tau_i &\leq \Jin + \frac{\Win}{\omega} \lambda_{\rm max}(\Jbb_i) \label{eq:torque_bound}.
\end{align}
\eqreftwo{force_bound}{torque_bound} are general bounds to the force or torque that can be exerted on any scatterer, given only the incident-field properties and the power and momentum properties of the relevant scattering channels. Intuitively, \eqref{force_bound} predicts an optimal force for nanoparticles that absorb all of the momentum along direction $i$ of the coupled incoming channels, and generate outgoing waves in those channels of equal power and large, negative momentum. The eigenvalue encodes the relative difficulty in any set of scattering channels of generating such momentum transfer. The analogous interpretation applies to \eqref{torque_bound} in terms of angular momentum.

Natural scattering channels for wavelength-scale nanoparticles are the vector spherical waves (VSWs), $\Mlm$ (TE) and $\Nlm$ (TM), where $\ell$ and $m$ are the angular and projected quantum numbers, respectively. A scatterer of finite size will have non-trivial coupling to only a finite number of channels parametrized by $\lmax$, a maximum angular quantum number. Farsund and Felderhof~\cite{Farsund1996} have derived analytical expressions for the integrals defining the matrices $\Pbb_i$ and $\Jbb_i$ (see \SM). As shown in \figref{fig1}, $\Jbb_z$ is diagonal, since the VSWs are pure angular momentum states. Conversely, $\Pbb_z$ has nonzero entries only off the diagonal. In the {\SM} we derive bounds on the largest eigenvalues of $\Pbb_i$ and $\Jbb_i$: $\lambda_{\rm max}(\Pbb_i) \leq 1$, and $\lambda_{\rm max}(\Jbb_i) = \lmax$. The physical origin of these bounds can be understood as follows: for any linear combination of VSWs comprising a single photon, they will demonstrate less directionality (and hence smaller linear momentum) than the corresponding plane-wave photon with momentum $\hbar \vect{k}$; by contrast, the angular momentum of a VSW can be as large as $\lmax$ times $\hbar |\vect{k}|$ per photon. We can simplify \eqreftwo{force_bound}{torque_bound} for prototypical plane-wave incident fields. Within channels up to $\lmax$, a plane wave with amplitude $\vect{E}_0$ and wavevector $\vect{k}$ carries power $\cin^\dagger \cin = \pi (\lmax^2 + 2\lmax) \frac{|\vect{E}_0|^2}{2 Z_0 |\vect{k}|^2}$ (\SM). Its linear momentum per time is $\Pin = \cin^\dagger \Pbb_i \cin / c = \frac{\beta_i}{c} \frac{\lmax}{\lmax+1} \cin^\dagger \cin$, where $\beta_i = \hat{\vect{k}} \cdot \hat{\vect{i}}$ is the fraction of the incident wave's momentum in direction $i$. Its angular momentum per time is $\Jin = \cin^\dagger \Jbb_i \cin /\omega = (\beta_i \gamma_i  / \omega) \cin^\dagger \cin$, where $\gamma_i$ is the degree of right circular polarization for the wave projected into direction $i$. Both $\beta_i$ and $\gamma_i$ have a range of [-1,1]. Following this procedure and dividing out the plane-wave intensity, $I_{\rm inc} = |\vect{E}_0|^2 / 2Z_0$, yields the bounds
\begin{align}
    \frac{F_i}{I_{\rm inc}} &\leq \frac{\lambda^2}{4\pi c} \left(\lmax^2 + 2\lmax\right) \left(1 + \beta_i \frac{\lmax}{\lmax+1}\right) \label{eq:force_bound_pw}, \\
    \frac{\tau_i}{I_{\rm inc}} &\leq \frac{\lambda^3}{8\pi^2 c} \left(\lmax^2 + 2\lmax\right)\left(\lmax + \beta_i\gamma_i\right). \label{eq:torque_bound_pw}
\end{align}
\Eqreftwo{force_bound_pw}{torque_bound_pw} bound the largest forces/torques that can be generated from incident plane waves. (\eqreftwo{force_bound}{torque_bound} provide bounds for more general incident waves.) The quantities $\lambda^2/c$ and $\lambda^3/c$ naturally emerge as force/torque analogs of the $\lambda^2$ scattering cross-sections. Such proportionalities emerge physically by dimensional analysis, while the quadratic framework leading to \eqreftwo{force_bound_pw}{torque_bound_pw} provides exact, quantitative upper bounds. 

Within \eqreftwo{force_bound_pw}{torque_bound_pw} is a second interesting result: spherically symmetric scatterers \emph{cannot} reach the plane-wave incident-field bounds, except in the trivial case $\lmax = 1$. Reaching these bounds requires outgoing waves to be proportional to the maximal eigenvectors of the $\Pbb_i$ and $\Jbb_i$ matrices, which do not coincide with the incoming-wave coefficients of a plane wave. (By modifying the incident field to match the VSW coefficients over the full $4\pi$ angular range, one can engineer a scenario in which spherically symmetric objects are optimal.) This is in contrast to scattered-power optimization, where it is known that spherically-symmetric scatterers can be globally optimal~\cite{Miroshnichenko2018} for \emph{any} incident field, and it arises because the additional requirements of directionality/polarization for linear/angular momentum require specific combinations of VSW channels for maximum effect.
\begin{figure}
    \centering
    \includegraphics[width=\linewidth]{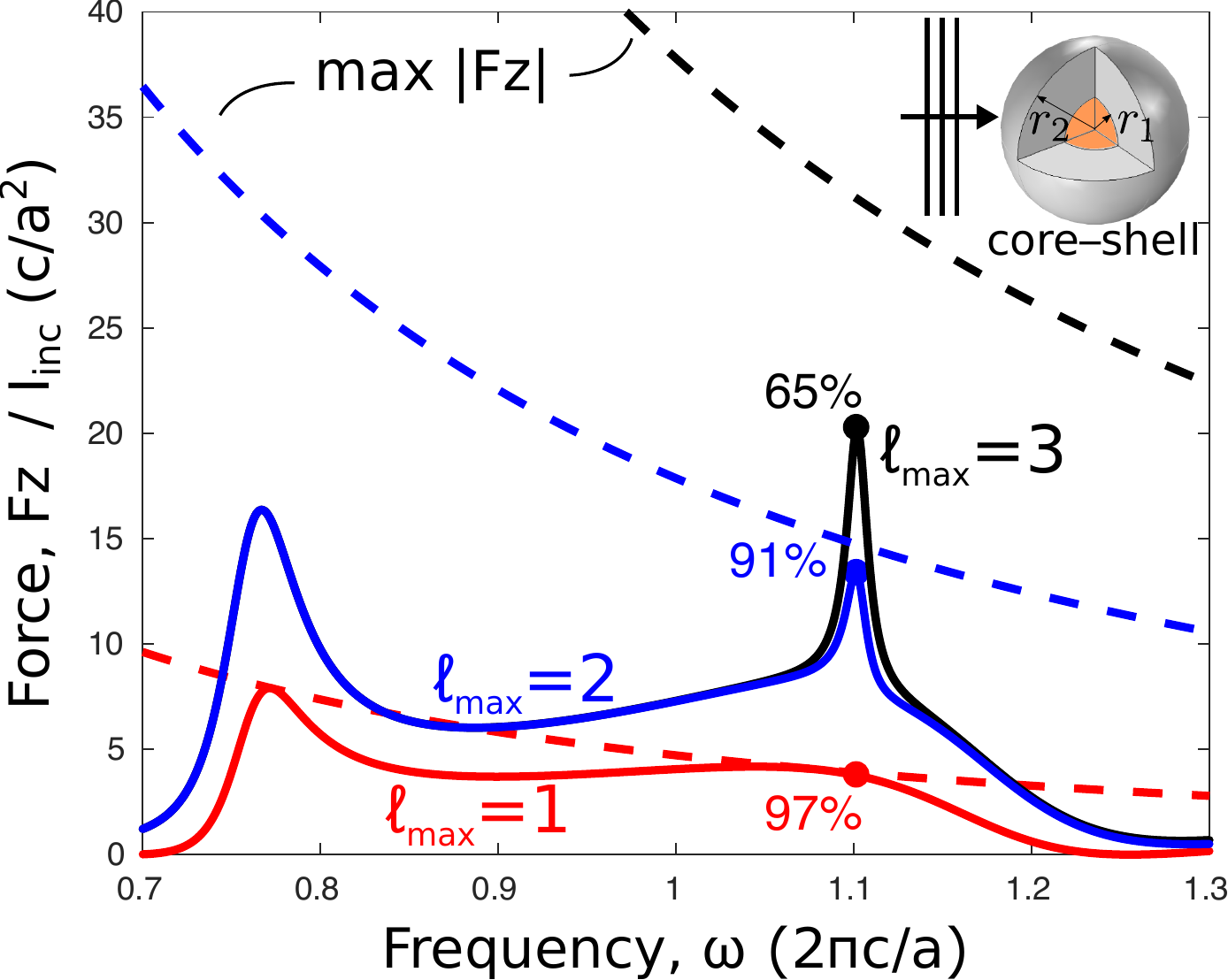}
    \caption{The force bounds of \eqreftwo{force_bound}{force_bound_pw} require strong and highly directional scattering. Core-shell structures with aligned resonances show strong scattering and imperfect but good directionality. Optimized Si--\ce{SiO_2} structures ($r_1 = 0.1a$, $r_2=0.9a$) experience a force approaching the $\lmax=3$ bounds, with negligible scattering in higher channels.}
    \label{fig:fig2}
\end{figure}

\begin{figure}
    \centering
    \includegraphics[width=\linewidth]{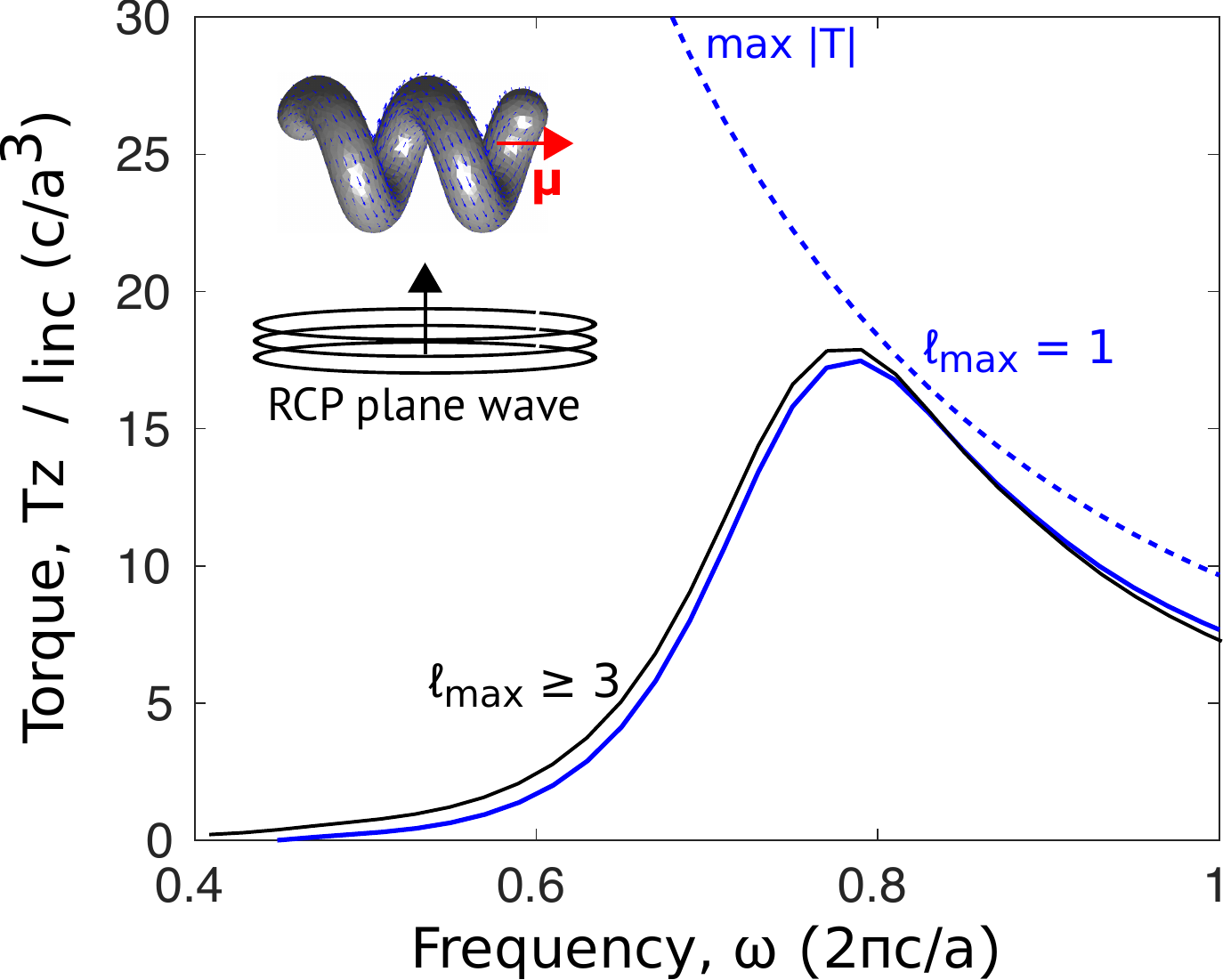}
    \caption{To achieve the torque bounds of \eqreftwo{torque_bound}{torque_bound_pw}, a scatterer must generate the largest possible $\Delta m$ between incoming/outgoing waves. A helix, interacting with a circularly polarized wave impinging normal to its rotation axis, exhibits a magnetic dipole moment $\vect{\mu}$ that generates counter-rotating outgoing waves, to create a torque (solid lines) that nearly achieves the $\lmax=1$ bound (dotted lines).}
    \label{fig:fig3}
\end{figure}

Figures~\ref{fig:fig2} and \ref{fig:fig3} show examples of designed nanoparticles that can approach the plane-wave bounds. For \figref{fig2}, the inner radii of core-shell Si--\ce{SiO_2} structures were optimized (over a length scale $a$) to exhibit aligned resonances (``super-scattering''~\cite{Ruan2011}). Even though spheres cannot exactly reach the bounds, as proven above, such resonances can effectively scatter light in the backwards direction, enhancing a large force in the forward direction that achieves a substantial fraction of the bound. For the three channels primarily excited, nearly 65\% of the total bound can be achieved, while nearly saturating the force due to the $\ell=1,2$ channels. By contrast, spheres cannot generate substantial torque, which requires coupling positive- and negative-angular-momentum channels. Helices are excellent nanoswimmers~\cite{Walker2015}, and we find that illuminating a helix (refractive index 3.5, structural details in \SM) \emph{normal} to its rotation axis generates counter-rotating outgoing waves and a large net torque perpendicular to its rotation axis. \Figref{fig3} shows that an optimized helix can closely approach the $\lmax = 1$ bound.

\begin{figure*}
    \centering
    \includegraphics[width=\linewidth]{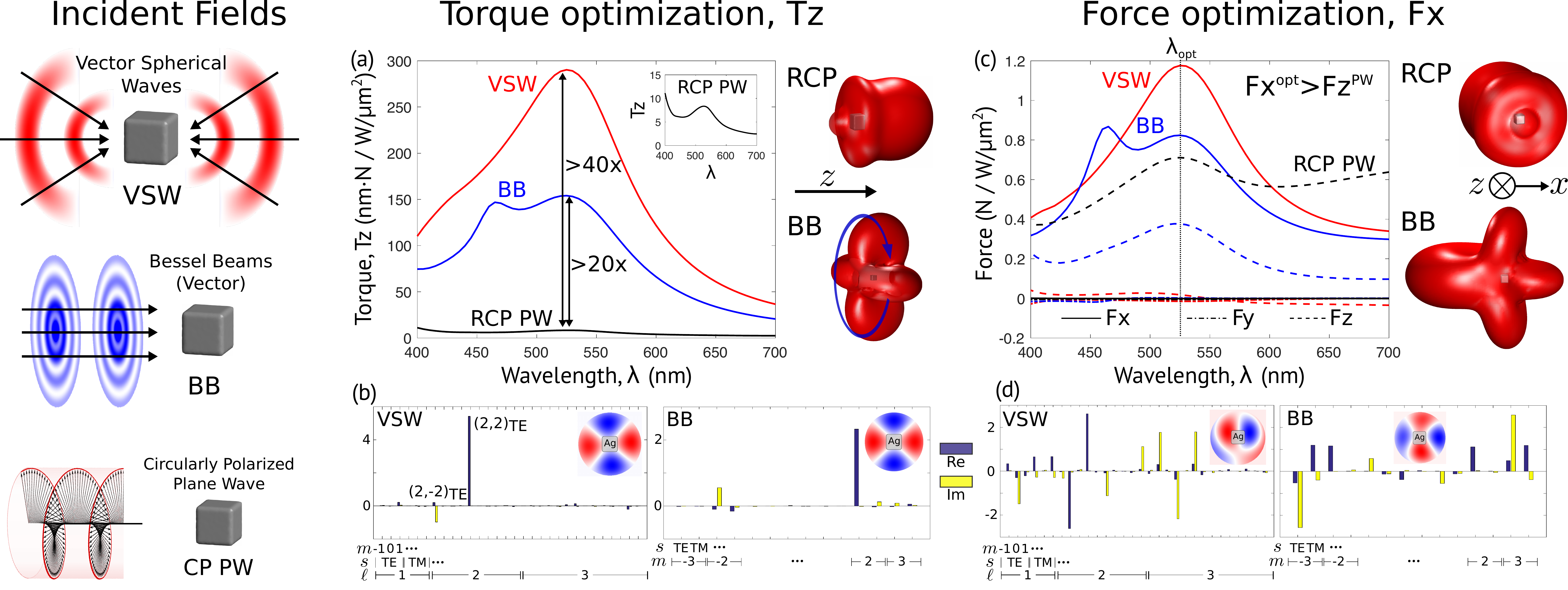}
    \caption{Global optimization of an illumination field can be achieved in a single eigenvector computation per \eqref{inc_field}. Here we optimize force and torque on a silver cube (\SI{200}{nm} edge length) for illumination fields decomposed into VSW and Bessel-beam (BB) bases, with circularly polarized plane waves (CP PWs) as a standard for comparison (left). (a) Despite the seemingly large torque generated on resonance ($\lambda = \SI{525}{nm}$) by a RCP PW (black line and inset), optimal VSW and BB incident fields offer $>$40$\times$ and $>$20$\times$ improvements, respectively, for a fixed field intensity. The scattered fields (right) for the optimal BB show an outgoing radiation pattern carrying angular momentum, primarily in the $\ell=2, m=\pm2$ channels. (c) Plane waves generate no \emph{in-plane} forces ($F_x$, $F_y$) on such a cube. VSW and BB incident fields optimized for maximum $|F_x|$ generate in-plane forces larger than the $F_z$ of a plane wave. The scattered fields (right) for the optimal BB show the highly asymmetric radiation pattern. (b,d) Optimized VSW and BB field coefficients, alongside field patterns in the plane of the cube (insets).}
    \label{fig:fig4}
\end{figure*}

\section*{Optimal illumination fields}
The quadratic framework lends itself readily to the reverse problem: given a fixed scatterer, what incident field generates maximal force/torque? More generally, what incident field maximizes general power/momentum quadratic forms? Significant interest in this problem has led to a variety of iterative optimization methods, which often converge to suboptimal local extrema~\cite{Lee2017}. Yet starting from the Poynting-vector/stress-tensor quadratic form, described by \eqref{quadform}, one can write any figure of merit in the form:
\begin{align}
    Q = \begin{pmatrix}
        \cin \\
        \cout
    \end{pmatrix}^\dagger
    \begin{pmatrix}
        \Qbb_{11} & \Qbb_{12} \\
        \Qbb_{12}^\dagger & \Qbb_{22}
    \end{pmatrix}
    \begin{pmatrix}
        \cin \\
        \cout
    \end{pmatrix},
    \label{eq:quadform2}
\end{align}
where $\Qbb_{11}$ and $\Qbb_{22}$ (and hence the whole $\Qbb$ matrix) are Hermitian. Per \eqrefrange{power}{torque}, for absorbed power, net force, and net torque, the off-diagonal terms are zero, while $\Qbb_{11} = -\Qbb_{22} = \mathbb{I}$, $\Pbb_i$, and $\Jbb_i$, respectively (where $\mathbb{I}$ is the identity matrix). For the power or momentum flux in the scattered field, the precise definitions of $\Qbb_{ii}$ depend on the decomposition of the incident field into incoming versus outgoing waves; in the VSW basis, one can show (cf. SM) that $\Qbb_{12} = -\Qbb_{11} = -\Qbb_{22} = \mathbb{I}$, $\Pbb_i$, or $\Jbb_i$, respectively. The outgoing-field coefficients are given by the product of the scattering matrix with the incoming-field coefficients, $\Sbb \cin$, such that one can rewrite \eqref{quadform2} as a quadratic form of the incoming-field coefficients only:
\begin{align}
    Q = \cin^\dagger \left[ \Qbb_{11} + \Qbb_{12} \Sbb + \Sbb^\dagger \Qbb_{12}^\dagger + \Sbb^\dagger \Qbb_{22} \Sbb \right] \cin.
    \label{eq:quadform3}
\end{align}
Constraining the total power contained in the incoming wave over some spatial region or set of channels imposes a constraint $\cin^\dagger \mathbb{A} \cin \leq 1$ for a Hermitian positive-definite matrix $\mathbb{A}$ (e.g., $\mathbb{A}$ is the identity matrix for a unity-average-power constraint in the scattering channels). The optimal coefficient vector $\cin{\rm (opt)}$ that maximizes \eqref{quadform3} subject to this constraint solves the generalized eigenproblem
\begin{align}
    \left[ \Qbb_{11} + 2 \Re (\Qbb_{12} \Sbb ) + \Sbb^\dagger \Qbb_{22} \Sbb \right] \cin^{\rm (opt)} = \lambda_{\rm max} \mathbb{A} \cin^{\rm (opt)},
    \label{eq:inc_field}
\end{align}
where $\lambda_{\max}$ is the largest eigenvalue. The extremal eigenfunction solving \eqref{inc_field} is the globally optimal incident field. Intuitively, it is sensible that the scattering matrix $\Sbb$ determines the optimal incident field, since $\Sbb$ encodes the response for any incoming wave. A key feature of \eqref{inc_field} is that for the wavelength-scale scatterers in many optical force experiments, only a small to moderate number of VSWs are typically excited. Hence $\Sbb$ has relatively few degrees of freedom, enabling rapid computation of the optimal incident field.

\Figref{fig4} demonstrates the capability for \eqref{inc_field} to generate orders-of-magnitude increases in force/torque through wavefront shaping. We consider a \SI{200}{nm} silver nanocube. The nanocube supports a strongly scattering quadrupole resonance at wavelength $\lambda = \SI{525}{nm}$ that already generates a significant force along the direction of an incoming plane wave. For a right circularly polarized (RCP) wave, absorption in the silver transfers the $m=1$ angular momentum of the wave to the cube and generates a commensurate torque (a, inset). Yet through wavefront shaping, the torque can be dramatically enhanced, without increasing the intensity of the incident field. We consider two incident-field bases: VSWs, with quantum numbers $\ell$, $m$, and $s$ (where $s$ denotes polarization), and vector Bessel beams (BBs)~\cite{Novitsky2007}, diffraction-free cylindrical beams with an angular order $m$ and a polarization $s$. Note that Bessel beams are a subset of VSWs, so VSWs can exhibit superior performance, though BBs are more practical for experimental implementations~\cite{Jones2015}. One could similarly optimize plane waves coming from within a given solid angle. After solving for the scattering matrix with a free-software implementation~\cite{ReidScuffEM} of the boundary element method~\cite{Harrington1993}, solution of \eqref{inc_field} yielded the optimal VSW and BB fields. As shown in \figref{fig4}(a,b), isolation and optimization of the dominant scattering channels yields 20--40$\times$ increases in the torque. The field patterns (right) indicate the angular momentum carried away by the scattered fields. In contrast to the torque case, the nanocube already feels large forces in plane-wave interactions, as seen in \figref{fig4}(c) (black dashed line), simply through the momentum carried in forward-scattered and backscattered waves (i.e., along $z$). However, the force in a lateral direction is necessarily zero by symmetry. With the same nanocube scattering matrix, we thus optimized \eqref{inc_field} for the $x$-directed force [\figref{fig4}(c)]. The optimal field coefficients, shown in \figref{fig4}(d), generate lateral forces even larger than the normally directed force under plane-wave excitation. The field patterns (right, red) show the highly asymmetric scattering that is responsible for the large lateral force. 

The quadratic-optimization approach developed here can be applied across the landscape of optical force and torque generation. The analytical bounds of \eqrefrange{force_bound}{torque_bound_pw} predict optimal response for a fixed incident field, while the optimal-eigenvector approach of \eqref{inc_field} determines optimal incident fields for a fixed structure. Looking forward, incorporation of temporal dynamics and associated effects (e.g., back-action~\cite{Juan2009}) may lead to robust and efficient methods for producing even larger effects, towards optimal dynamical control at the nanoscale.

\begin{acknowledgements} 
The authors thank Chia-Wei Hsu and Ognjen Ilic for helpful discussions. Y.L. and O.D.M. were supported by the Air Force Office of Scientific Research under award number FA9550-17-1-0093. L.F. was supported by a Shanyuan Overseas scholarship from the Hong Kong Shanyuan foundation at Nanjing University. S.G.J was supported in part by the Army Research Office under contract number W911NF-13-D-0001. N.F. was supported by the Air Force Office of Scientific Research (AFOSR) Multidisciplinary Research Program of the University Research Initiative (MURI), and from KAUST-MIT agreement \#2950. 
\end{acknowledgements}

\bibliography{library}

\end{document}


\preprint{APS/123-QED}

\title{Supporting Information\\Optimal nanoparticle forces, torques, and illumination fields}

\author{Yuxiang Liu}
\affiliation{Department of Applied Physics and Energy Sciences Institute, Yale University, New Haven, CT 06511}
\author{Lingling Fan}
\affiliation{Department of Applied Physics and Energy Sciences Institute, Yale University, New Haven, CT 06511}
\affiliation{School of Physics and National Laboratory of Solid State Microstructures, Nanjing University, Nanjing 210093, China}
\author{Yoonkyung E. Lee}
\affiliation{Department of Mechanical Engineering, Massachusetts Institute of Technology, Cambridge, MA 02139}
\author{Nicholas X. Fang}
\affiliation{Department of Mechanical Engineering, Massachusetts Institute of Technology, Cambridge, MA 02139}
\author{Steven G. Johnson}
\affiliation{Departments of Mathematics and Physics, Massachusetts Institute of Technology, Cambridge, MA 02139}
\author{Owen D. Miller}
\affiliation{Department of Applied Physics and Energy Sciences Institute, Yale University, New Haven, CT 06511}

\pagestyle{fancy}
\fancyhf{}
\fancyfoot[R]{\thepage}
\renewcommand{\headrulewidth}{0pt}

\date{\today}

\maketitle
\thispagestyle{fancy}

\tableofcontents

\section{Time-reversal properties of energy/momentum flux operators}
In the main text, we saw that energy/momentum quantities of interest, such as absorbed power, force, or torque, can generally be written for a field $\psi$ as quadratic forms
\begin{align}
    Q = \int_S \psi^\dagger \mathcal{Q} \psi,
    \label{eq:quadform}
\end{align}
where $\mathcal{Q}$ is a Hermitian operator determined by the Poyting vector or the electromagnetic stress tensor. Here, we use the fact that $\mathcal{Q}$ represents energy or momentum flow to assert that $\mathcal{Q}$ must satisfy a general time-reversal expression.

Consider a total field that is purely outgoing: $\psi = \Vout \cv$. Then $Q$ would be given by
\begin{align}
    Q = \cv^\dagger \left( \int_S \Vout^\dagger \mathcal{Q} \Vout \right) \cv
    \label{eq:Q}
\end{align}
If we time-reverse the fields, $\Vout \rightarrow \Vin = \mathcal{P} \cc{\Vout}$, then the quantity $Q$ must go to its negative (energy/momentum flows in the opposite direction):
\begin{align}
    -Q &= \cv^\dagger \left( \int_S \Vout^T \mathcal{P} \mathcal{Q} \mathcal{P} \cc{\Vout} \right) \cv \nonumber \\
       &= \cv^\dagger \left( \int_S \Vout^\dagger \mathcal{P} \mathcal{Q}^T \mathcal{P} \Vout \right) \cv.
       \label{eq:Qtrans}
\end{align}
Since \eqref{Q} and \eqref{Qtrans} apply for any $\cv$ and any $\Vbb_{+/-}$, we have the relation
\begin{empheq}[box=\widefbox]{align}
    \mathcal{Q} = -\mathcal{P} \mathcal{Q}^T \mathcal{P}.
    \label{eq:Q_TR}
\end{empheq}
From \eqref{Q_TR}, it is straightforward to show that, as argued in the main text, that the incoming/outgoing channels carry equal and opposite energy/momentum:
\begin{empheq}[box=\widefbox]{align}
    \int_S \Vout^\dagger \mathcal{Q} \Vout = - \int_S \Vin^\dagger \mathcal{Q} \Vin.
    \label{eq:VQV}
\end{empheq}


\section{Quadratic forms}
\label{sec:Quad}

First, we show that we can write the flux rates of power, linear momentum, and angular momentum through any surface $S$ as the quadratic form given by \eqref{quadform} and repeated here,
\begin{align}
    Q = \int_S \psi^\dagger \mathcal{Q} \psi,
    \label{eq:quadform2}
\end{align}


\subsection{Power}
Assuming an outward normal $\nhat$ on some surface $S$, net power flow in a field $\psi$ is given by the Poynting vector, which can be written in six-vector notation as
\begin{empheq}[box=\widefbox]{align}
    P &= \int_S \psi^\dagger \left(-\frac{1}{4} \Theta\right) \psi,
\end{empheq}
where $\Theta$ is the real-symmetric matrix,
\begin{align}
    \Theta = \begin{pmatrix} & -\nhat \times \\ \nhat \times & \end{pmatrix}.
\end{align}

\subsection{Linear momentum}
The flux of linear momentum through a surface is determined by a surface integral of the Maxwell stress tensor, which is $\sigmat = \left[\Ev \Ev^\dagger - \frac{1}{2}\bm{\mathcal{I}} \left(\Ev^\dagger \Ev\right) \right] + \left[\Hv \Hv^\dagger - \frac{1}{2} \bm{\mathcal{I}} \left(\Hv^\dagger \Hv\right)\right]$ (for $\epso = \mu_0 = 1$). The linear-momentum flux along a given direction, denoted $\xhat$, is given by
\begin{align}
    \vect{P} \cdot \xhat = \frac{1}{2} \Re \int_S \xhat \cdot \left\lbrace \left[\Ev \Ev^\dagger - \frac{1}{2}\bm{\mathcal{I}} \left(\Ev^\dagger \Ev\right) \right] + \left[\Hv \Hv^\dagger - \frac{1}{2} \bm{\mathcal{I}} \left(\Hv^\dagger \Hv\right)\right] \right\rbrace \nhat.
\end{align}
If we define the nonsquare, $6\times2$ matrices $\mathbb{X}$ and $\mathbb{N}$,
\begin{align}
    \mathbb{N} = 
    \begin{pmatrix}
        \nhat & \\
        & \nhat
    \end{pmatrix}, \qquad
    \mathbb{X} = 
    \begin{pmatrix}
        \xhat & \\
        & \xhat
    \end{pmatrix},
\end{align}
then we can alternatively write the flux rate as
\begin{align}
    \vect{P} \cdot \xhat = \frac{1}{2} \Re \int_S \left[ \Tr \left( \mathbb{X}^T \psi \psi^\dagger \mathbb{N} \right) - \frac{1}{4} \psi^\dagger \psi \Tr \left(\mathbb{X}^T \mathbb{N}\right) \right].
\end{align}
By straightforward trace manipulations, we can rewrite this as
\begin{empheq}[box=\widefbox]{align}
    \vect{P} \cdot \xhat = \frac{1}{4} \int_S \psi^\dagger \left[ \mathbb{N} \mathbb{X}^T + \mathbb{X} \mathbb{N}^T - \frac{1}{2} \Tr\left(\mathbb{X}^T \mathbb{N}\right) \right] \psi,
    \label{eq:force_qf}
\end{empheq}
which is precisely of the form of \eqref{quadform2}, with one-fourth times the term in square brackets denoting the operator $\mathcal{Q}$.

\subsection{Angular momentum}
The angular-momentum integrand is similar to that for linear momentum, with the replacement $\sigmat \rightarrow \vect{r} \times \sigmat$. In the direction $\xhat$, the angular momentum (around the origin) takes the form
\begin{align}
    \vect{J} \cdot \xhat = \frac{1}{2} \Re \int_S \left(\xhat \times \vect{r}\right) \cdot \left\lbrace \left[\Ev \Ev^\dagger - \frac{1}{2}\bm{\mathcal{I}} \left(\Ev^\dagger \Ev\right) \right] + \left[\Hv \Hv^\dagger - \frac{1}{2} \bm{\mathcal{I}} \left(\Hv^\dagger \Hv\right)\right] \right\rbrace \nhat.
\end{align}
Clearly the angular-momentum flux is identical to the linear-momentum flux, with the replacement $\vect{x} \rightarrow \vect{r} \times \vect{x}$. Thus, if we define 
\begin{align}
    \mathbb{U} = 
    \begin{pmatrix}
        \vect{r} \times \vect{x} & \\
        & \vect{r} \times \vect{x}
    \end{pmatrix},
\end{align}
we can directly write the angular-momentum analog of \eqref{force_qf}:
\begin{empheq}[box=\widefbox]{align}
    \vect{J} \cdot \xhat = \frac{1}{4} \int_S \psi^\dagger \left[ \mathbb{N} \mathbb{U}^T + \mathbb{U} \mathbb{N}^T - \frac{1}{2} \Tr\left(\mathbb{U}^T \mathbb{N}\right) \right] \psi,
    \label{eq:torque_qf}
\end{empheq}
again with one-fourth times the term in square brackets denoting $\mathcal{Q}$.

\subsection{Power: scattering-coefficient quadratic forms}
Now we consider a scattering problem in which an incident field interacts with a scattering body, thereby producing a scattered field. Any field $\psi$ (which could be the total field, the scattered field, or the incident field, e.g.) can be decomposed into incoming- and outgoing-wave components, as in the main text, 
\begin{align}
    \psi &= \psiin + \psiout \nonumber \\
    &= \Vin \cin + \Vout \cout.
\end{align}
Then the power in $\psi$ flowing through $S$ is given by
\begin{align}
    Q &= \int_S \left[ \psiin^\dagger \mathcal{Q} \psiin + \psiout^\dagger \mathcal{Q} \psiout + 2 \Re \psiin^\dagger \mathcal{Q} \psiout \right] \nonumber \\
    &= \cin^\dagger \left( \int_S \Vin^\dagger \mathcal{Q} \Vin \right) \cin + \cout^\dagger \left( \int_S \Vout^\dagger \mathcal{Q} \Vout \right) \cout + 2\Re\left[\cin^\dagger \left( \int_S \Vin^\dagger \mathcal{Q} \Vout \right) \cout \right],
    \label{eq:Qall}
\end{align}
where $\mathcal{Q}$ is the corresponding power/momentum operator from the previous subsections. In the main text, we saw that we time-reversal incoming/outgoing basis states can be chosen to satisfy $\int_S \Vin^\dagger \mathcal{Q} \Vin = -\int_S \Vout^\dagger \mathcal{Q} \Vout$ and $\int_S \Vin^\dagger \mathcal{Q} \Vout = 0$, giving
\begin{align}
    Q &= \cin^\dagger \Qbb_{\rm in} \cin - \cout^\dagger \Qbb_{\rm in} \cout.
    \label{eq:Qinout}
\end{align}
where $\Qbb_{\rm in} = \int_S \Vin^\dagger \mathcal{Q} \Vin$.

Absorbed power is simply the power flow of the total field into $S$, and thus can be written identically from \eqref{Qinout}, where $\cin$ and $\cout$ now refer specifically to the in/out decomposition of the \emph{total} field. Moreover, for power flow, as discussed in the main text, it is convenient to choose $\Qbb_{\rm in} = \mathbb{I}$, where $\mathbb{I}$ is the identity matrix, such that
\begin{empheq}[box=\widefbox]{align}
    \Pabs = \cin^\dagger \cin - \cout^\dagger \cout.
    \label{eq:pabs}
\end{empheq}
Scattered power is the \emph{outgoing} power in the \emph{scattered} field, which has no incoming-field component and can thus be written $\Pscat = \cscat^\dagger \cscat$. Different bases may have different partitions for the incident/scattered fields in the in/out basis; for vector spherical waves, 
\begin{align}
    \vect{c}_{\rm in} &=  \frac{1}{2}\vect{c}_{\rm inc} ~,  \label{eq:in}\\
    \vect{c}_{\rm out}&=\vect{c}_{\rm scat} + \frac{1}{2} \vect{c}_{\rm inc}~. \label{eq:out} 
\end{align}
Thus, 
\begin{empheq}[box=\widefbox]{align}
    \Pscat = \left(\cout - \cin\right)^\dagger \left(\cout - \cin\right).
    \label{eq:pscat}
\end{empheq}
Extinction is the sum of absorption and scattering, and thus in the VSW basis is the sum of \eqref{pabs} and \eqref{pscat}, giving
\begin{empheq}[box=\widefbox]{align}
    \Pext = 2 \Re \left[ \cin^\dagger \left( \cin - \cout \right) \right]
    \label{eq:pext}
\end{empheq}

\subsection{Force/torque: scattering-coefficient quadratic forms}
We can work out similar quadratic forms, in terms of the scattering-channel coefficients, for the force, torque, and scattering/extinction contributions to the corresponding momentum flux rates. \Eqref{Qinout} holds for any of these quadratic forms, beyond just power. Force is the net transfer of linear momentum in the total field $\psi$, and thus by analogy with \eqref{pabs} (but noting that in this case the corresponding matrix is not the identity):
\begin{empheq}[box=\widefbox]{align}
    \vect{F} \cdot \xhat = \cin^\dagger \Pbb_i \cin - \cout^\dagger \Pbb_i \cout,
    \label{eq:force}
\end{empheq}
where $\Pbb_i = \int_S \Vin^\dagger \mathcal{Q} \Vin$ for $\mathcal{Q}$ as defined by \eqref{force_qf}.

Then, the linear-momentum flux rate for the scattered field, $\psi_{\rm scat}$, is given by the same expression, except with no incoming-wave component, the sign of the outgoing-wave component reversed, and the outgoing-wave coefficients replaced with the scattered-field coefficients: $\vect{P}_{\rm scat} \cdot \xhat = \cscat^\dagger \Pbb_i \cscat$. In the VSW basis, by \eqreftwo{in}{out},
\begin{empheq}[box=\widefbox]{align}
    \vect{P}_{\rm scat} \cdot \xhat = \left(\cout - \cin\right)^\dagger \Pbb_i \left(\cout - \cin\right).
    \label{eq:fscat}
\end{empheq}
Then, the linear momentum extinguished is
\begin{empheq}[box=\widefbox]{align}
    \vect{P}_{\rm scat} \cdot \xhat = 2 \Re \left[ \cin^\dagger \Pbb_i \left(\cout - \cin\right) \right].
    \label{eq:fext}
\end{empheq}

For angular momentum, the corresponding equations take the same form as \eqrefrange{force}{fext}, with the replacement $\Pbb_i \rightarrow \Qbb_i$. In the next section, we list the definitions of vector-spherical-waves and use the results of \citeasnoun{Farsund1996} to explicitly write out the matrices $\Pbb_i$ and $\Qbb_i$.

\section{Vector spherical waves: definitions and matrices}
\label{sec:VSW}
There are many possible conventions for vector spherical waves (VSWs), with different coefficient and sign conventions, and thus for clarity we include our convention here in detail (our convention is the same as that of \citeasnoun{Tsang2000}), and we also include the force and torque matrices $\Pbb_i$ and $\Jbb_i$ in the VSW basis. 

First, we note that in addition to the in/out basis used throughout, one could instead use an incident-field/scattered-field separation. Which separation is used determines which types of spherical Bessel functions are used in the VSWs:
\begin{subequations}
\begin{align}
\Ev_{\rm inc}(\xv) &= \Vbb^{\rm reg}(\xv) \vect{c}_{\rm inc} \label{eq:Einc} \\
\Ev_{\rm scat}(\xv) &= \mathbb{V}^{+}(\xv) \vect{c}_{\rm scat} \label{eq:Escat}\\
\Ev_{\rm in}(\xv) &= \mathbb{V}^{-}(\xv) \vect{c}_{\rm in} \label{eq:Ein}  \\
\Ev_{\rm out}(\xv) &= \mathbb{V}^{+}(\xv) \vect{c}_{\rm out} \label{eq:Eout},
\end{align}
\end{subequations}
where the ``reg'' subscript denotes ``regular'' (i.e. well-behaved spherical Bessel functions at the origin), the ``+'' superscript denotes outgoing waves, and the ``-'' superscript denotes incoming waves.  (In the main text it was clearer to use subscripts, to avoid conjugate-transpose symbol clashes, but here we use superscripts to avoid index symbol clashes.) The tensors $\mathbb{V}^{\rm reg/+/-}$ comprise the vector spherical waves as columns:
\begin{align}
    \mathbb{V}^{\rm reg/+/-}(\xv) &= \begin{bmatrix} \ldots  \vect{N}^{\rm reg/+/-}_{\ell,m}(\xv) ~, \quad \vect{M}^{\rm reg/+/-}_{\ell,m}(\xv)  \ldots \end{bmatrix}~, \nonumber \\
 1 &\leq \ell \leq \ell_{\rm max}~, -\ell \leq m \leq \ell~.
    \end{align} 
The vectors $\vect{N}^{\rm reg/+/-}_{\ell,m}(\xv)$ denote $e$-polarized waves while $\vect{M}^{\rm reg/+/-}_{\ell,m}(\xv)$ denote $h$-polarized waves. $\ell$ is the angular momentum ``quantum number'' while $m$ is the projected (angular momentum) quantum number. The magnetic fields are given by the same equations as the electric fields, with $\vect{M} \rightarrow \vect{N}$ and $\vect{N} \rightarrow -\vect{M}$.

Our vector-spherical-wave convention is
\begin{align}
\vect{N}^{\rm reg/+/-}_{\ell,m}(\xv) &=  \frac{1}{k} \nabla \times 
 \left[ \nabla \times \left(\xv ~ z_\ell^{\rm reg/+/-}(kr)Y_{\ell m} (\theta, \phi) \right) \right] \label{eq:MN1}\\
\vect{M}^{\rm reg/+/-}_{\ell,m}(\xv) &=   \nabla \times \left(\xv ~ z_\ell^{\rm reg/+/-}(kr)Y_{\ell m}
 (\theta, \phi) \right) \label{eq:MN2}~,
\end{align}
where $z_\ell^{\rm reg/+/-}$ represents the three spherical Bessel functions $j_\ell, h_\ell^{(1)}$ and $h_\ell^{(2)}$ respectively (also see \cite[Eqs.~(4.9)--(4.14)]{Bohren1983}). The spherical harmonics $Y_{\ell m}$ are defined as 
\begin{equation}
\label{eq:SH}
Y_{\ell m}(\theta, \phi) = \sqrt{\frac{2\ell+1}{4\pi \ell (\ell+1)} \frac{(\ell-m)!}{(\ell+m)!}} P_\ell^m(\cos \theta) e^{im\phi}~.
\end{equation}
Our definition of the vector spherical waves are the same as that in \cite[Eqs.~(1.4.56,1.4.57)]{Tsang2000}. Note that the spherical harmonics defined in \eqref{SH} for different $\ell$'s and $m$'s are orthogonal but not unit-normalized, as 
\begin{equation}
\int Y_{\ell m}(\theta, \phi)^*Y_{\ell' m'}(\theta, \phi) = \frac{1}{\ell (\ell +1)} \delta_{\ell \ell'} \delta_{m m'}~.
\end{equation}

Applying the curl operator in \eqreftwo{MN1}{MN2} becomes \cite{Bohren1983}
\begin{align}
\vect{N}^{\rm reg/+/-}_{\ell,m}(\xv) =&  \sqrt{\frac{2\ell+1}{4\pi \ell (\ell+1)} \frac{(\ell-m)!}{(\ell+m)!}} \cdot 
\left( \frac{z^{\rm reg/+/-}_\ell (\rho)}{\rho}e^{im\phi} \ell (\ell+1)P_\ell^m (\cos\theta) \hat{\vect{e}}_r \right. \nonumber \\
 &+ \left. e^{im\phi}\frac{dP_\ell^{m} (\cos\theta)}{d\theta}\frac{1}{\rho}\frac{d}{d\rho}\left[ \rho z_\ell^{\rm reg/+/-}(\rho)\right] \hat{\vect{e}}_\theta \right. \nonumber\\
 &+ \left. (im) e^{im\phi} \frac{P_\ell^{m} (\cos\theta)}{\sin\theta} \frac{1}{\rho}\frac{d}{d\rho}\left[ \rho z_\ell^{\rm reg/+/-}(\rho)\right] \hat{\vect{e}}_\phi\right)~, \label{eq:MN1exp}\\
\vect{M}^{\rm reg/+/-}_{\ell,m}(\xv) =&  \sqrt{\frac{2\ell+1}{4\pi \ell (\ell+1)} \frac{(\ell-m)!}{(\ell+m)!}} \cdot 
\left( (im)\frac{e^{im\phi} }{\sin\theta} P_\ell^m (\cos \theta)z_\ell^{\rm reg/+/-}(\rho) \hat{\vect{e}}_\theta \right. \nonumber\\
 &- \left. e^{im\phi}\frac{dP_\ell^{m} (\cos\theta)}{d\theta}z_\ell^{\rm reg/+/-}(\rho) \hat{\vect{e}}_\phi\right)~, \label{eq:MN2exp}
\end{align}
where $\rho = kr$. 

As we will discuss in the next subsection, Farsund and Felderhof~\cite{Farsund1996} worked out overlap integrals of the Maxwell stress tensor for vector spherical waves of different orders, which determine the values of the force and torque matrices whose eigenvalues we bound. We use a slightly different VSW convention from Farsund and Felderhof, which we delineate here:
\begin{enumerate}
\item In \citeasnoun{Farsund1996}, they define $Y_{\ell m}(\theta, \phi)$ to be 
$$
Y_{\ell m}(\theta, \phi) = \sqrt{\frac{2\ell+1}{4\pi} \frac{(\ell-m)!}{(\ell+m)!}} P_\ell^m(\cos \theta) e^{im\phi}~.
$$
In this definition, $Y_{\ell m}(\theta, \phi)$ is orthonormal. Therefore, we have a factor $\sqrt{\ell(\ell+1)}$ difference. 
\item Their definition of $\mathbb V^{\rm reg/+/-}$ has an extra factor $k$.
\item Their definition of $\mathbf M^{\rm reg/+/-}$ has an extra factor $i$.

Therefore, the conversion between our coefficients $\vect{c}$ and the Farsund--Felderhof coefficients $\vect{c}^{\rm FF}$ is 
\begin{align}
    \vect{c}_{e\ell m} &=  k \sqrt{\ell(\ell+1)}  \vect{c}_{e\ell m}^{\rm FF}, \\
    \vect{c}_{h\ell m} &=  i k  \sqrt{\ell(\ell+1)}  \vect{c}_{h\ell m}^{\rm FF}.
\end{align}
\end{enumerate}

\subsection{Torque matrices}
\label{sec:torque_VSW}
As shown in \secref{Quad}, the matrices $\Pbb_i$ and $\Jbb_i$, for force and torque in the $i$ direction, respectively, are determined by overlap integrals $\int_S \Vbb_-^\dagger \mathcal{Q} \Vbb_-$, involving the basis tensor $\Vbb_-$ and a tensor $\mathcal{Q}$ defined by the particular integral quantity (stress tensor, Poynting flow, etc.). In this subsection we write out the torque matrix $\Jbb_i$ (translating the results of \citeasnoun{Farsund1996}), while the next subsection contains the force matrix $\Pbb_i$.

The torque matrix $\Jbb_i$ accounts for nonzero integrals (over the spherical bounding surface) of VSWs of order $\{\ell,m,s\}$ with VSWs of order $\{\ell',m',s'\}$. Farsund and Felderhof show that it is simpler to work with a variable $q \in \{0,\pm 1\}$ instead of $i$, where $q=0$ corresponds to $i=z$ and $q = \pm 1$ are linear combinations of the $x$ and $y$ directions. For a given $q$, it is helpful to define a term $L_q(\ell m m')$ as follows:
$$
L_q(\ell mm') = (-1) ^{\ell+m+1} \sqrt{\ell(\ell+1)(2\ell+1)} \tj{\ell}{\ell}{1}{-m}{m'}{q}~, \quad q \in \{ -1, 0, 1 \}~,
$$
where the last term of the above equation is the Wigner-3j symbol~\cite{Farsund1996}. For any $q$ (and $i$), the torque matrix is block-diagonal in $\ell$, as there is no coupling between $\ell$ and $\ell'$ waves when $\ell \neq \ell'$. In terms of $L_q$, the $\ell$ blocks of the torque matrices are:
\begin{align}
\mathbb J_z ^\ell (mm') &= L_0(\ell mm') \\
\mathbb J_x ^\ell (mm')&= \frac{L_{+1}(\ell mm') - L_{-1}(\ell mm')}{\sqrt{2}} \\
\mathbb J_y ^\ell (mm')&= -i\frac{L_{+1}(\ell mm') + L_{-1}(\ell mm')}{\sqrt{2}}
\end{align}
Now, we want to write down the matrices $\mathbb J_x^\ell, \mathbb J_y^\ell$ and $\mathbb J_z^\ell$ explicitly and try to get the eigenvalues analytically. First, we have 
\begin{align}
L_0(\ell mm') &= (-1)^{\ell+m+1}\sqrt{\ell(\ell+1)(2\ell+1)} \tj{\ell}{\ell}{1}{-m}{m'}{0} \\
                &= (-1)^{\ell+m+1}\sqrt{\ell(\ell+1)(2\ell+1)}\cdot  (-1)^{\ell+m+1} \frac{m}{\sqrt{\ell(\ell+1)(2\ell+1)}}\delta_{mm'} \\
                &= m~\delta_{mm'}
\end{align}
Therefore we have 
\begin{equation}
\mathbb J_z^\ell = 
\begin{bmatrix}
    -\ell & & & & & &  \\
    & -\ell+1 &&&&& \\
    &&\ddots&&&&   \\
    &&&0&&& \\
    &&&&\ddots&& \\
    &&&&&\ell-1&  \\
    &&&&&&\ell
\end{bmatrix}
\end{equation}
It is clear that the eigenvalues of $J_z^\ell$ are $-\ell, -\ell+1, \hdots, 0, \hdots, \ell-1, \ell$.

\begin{align}
L_1(\ell mm') &= (-1)^{\ell+m+1}\sqrt{\ell(\ell+1)(2\ell+1)} \tj{\ell}{\ell}{1}{-m}{m'}{1} \\
                &= (-1)^{\ell+m+1}\sqrt{\ell(\ell+1)(2\ell+1)}\cdot  (-1)^{\ell+m} \sqrt{\frac{(\ell+m)(\ell-m+1)}{2\ell(\ell+1)(2\ell+1)}}\delta_{m',m-1} \\
                &= -\sqrt{\frac{(\ell+m)(\ell-m+1)}{2}}\delta_{m',m-1} 
\end{align}
If we want to write it explicitly, it is 
\begin{equation}
L_{1}^{\ell} = 
\begin{bmatrix}
    0 & 0& &0 &0   \\
    -\sqrt{\frac{1\cdot2\ell}{2}}&0  &&0&0\\
    &-\sqrt{\frac{2\cdot(2\ell-1)}{2}}&& 0&0   \\
    &&\ddots&& \\
    &&&-\sqrt{\frac{(2\ell-1)\cdot2}{2}} &0  \\
    &&&&-\sqrt{\frac{2\ell\cdot1}{2}} 
\end{bmatrix}
\end{equation}

\begin{align}
L_{-1}(\ell mm') &= (-1)^{\ell+m+1}\sqrt{\ell(\ell+1)(2\ell+1)} \tj{\ell}{\ell}{1}{-m}{m'}{-1} \\
                &= (-1)^{\ell+m+1}\sqrt{\ell(\ell+1)(2\ell+1)}\cdot  (-1)^{\ell+m+1} \sqrt{\frac{(\ell-m)(\ell+m+1)}{2\ell(\ell+1)(2\ell+1)}}\delta_{m',m+1} \\
                &= \sqrt{\frac{(\ell-m)(\ell+m+1)}{2}}\delta_{m',m+1} 
\end{align}
If we want to write it explicitly, it is 
\begin{equation}
L_{-1}^{\ell} = 
\begin{bmatrix}
    0 & \sqrt{\frac{1 \cdot 2\ell}{2}}&0 & &0 &0  \\
     0 &0  &  \sqrt{\frac{2\cdot (2\ell-1)}{2}}&&0 &0\\
    & && \ddots&  & \\
    && & &  \sqrt{\frac{(2\ell-1)\cdot 2}{2}} &0 \\
    0&0&0&&0&\sqrt{\frac{2\ell\cdot1}{2}} 
\end{bmatrix}
\end{equation}
We have 
\begin{align}
\mathbb J_x^\ell = \begin{bmatrix}
    0 &\frac{\sqrt{1\cdot 2\ell}}{2} & & & & &  \\
    \frac{\sqrt{1\cdot 2\ell}}{2}& 0 & \frac{\sqrt{2\cdot (2\ell-1)}}{2}&&&& \\
     & \frac{\sqrt{2\cdot (2\ell-1)}}{2} &0 & \frac{\sqrt{3\cdot (2\ell-2)}}{2}&&&   \\
    &&\ddots&\ddots&\ddots&& \\
    &&&\frac{\sqrt{(2\ell-2)\cdot 3}}{2}&0&\frac{\sqrt{(2\ell-1)\cdot 2}}{2}&  \\
    &&&&\frac{\sqrt{(2\ell-1)\cdot 2}}{2}&0& \frac{\sqrt{2\ell\cdot 1}}{2}  \\
    &&&&&\frac{\sqrt{2\ell\cdot 1}}{2} &0
\end{bmatrix} \nonumber
\end{align}

\begin{align}
\mathbb J_y^\ell = \begin{bmatrix}
    0 &i\frac{\sqrt{1\cdot 2\ell}}{2} & & & & &  \\
    -i\frac{\sqrt{1\cdot 2\ell}}{2}& 0 & i\frac{\sqrt{2\cdot (2\ell-1)}}{2}&&&& \\
     & -i\frac{\sqrt{2\cdot (2\ell-1)}}{2} &0 & i\frac{\sqrt{3\cdot (2\ell-2)}}{2}&&&   \\
    &&\ddots&\ddots&\ddots&& \\
    &&&-i\frac{\sqrt{(2\ell-2)\cdot 3}}{2}&0&i\frac{\sqrt{(2\ell-1)\cdot 2}}{2}&  \\
    &&&&-i\frac{\sqrt{(2\ell-1)\cdot 2}}{2}&0& i\frac{\sqrt{2\ell\cdot 1}}{2}  \\
    &&&&&-i\frac{\sqrt{2\ell\cdot 1}}{2} &0
\end{bmatrix} \nonumber 
\end{align}
Although it is not analytically obvious how to derive the eigenvalues of $\Jbb_x^\ell$ or $\Jbb_y^\ell$, it is straightforward to show numerically that their eigenvalues are also $-\ell, -\ell+1, \hdots, \ell-1, \ell$. This can also be argued by symmetry: absent an incident field, there is no preferred direction in space, and thus the angular momentum ``available'' in any direction should be identical.
    
\subsection{Force matrices}
\label{sec:force_VSW}
The force matrices are more complex than the torque matrices. For any $q$, we can decompose the force matrices into two parts:
$$
\mathbb P_q = \mathbb P_q ^d  + \mathbb P_q ^c~, \quad q \in  \{ -1, 0, 1 \}~,
$$
where $\mathbb P_q ^d$ denotes the interaction for waves of the same polarization, but different $\ell$'s, whereas $\mathbb P_q ^c$ is the interaction matrix for the same $\ell$'s but different polarizations ($e$-$h$).

Again, following \citeasnoun{Farsund1996} while noting the different normalizations, 
\begin{equation} \label{eq:Fd}
\mathbb P_q^{d'}(\ell m,\ell'm') = \Re\left[ \sqrt{\frac{1}{\ell(\ell+1)}}\cdot  i^{\ell-\ell'} (\ell'^2 + \ell' -1)R_q(\ell m, \ell'm')  \cdot  \sqrt{\frac{1}{\ell'(\ell'+1)}} \right]~,
\end{equation}
where the term $R_q(\ell m, \ell'm')$ is a product of two Wigner 3j-symbols,
\begin{align}
R_q(\ell m,\ell'm') = (-1) ^{m} \sqrt{(2\ell+1)(2\ell'+1)} \tj{\ell}{\ell'}{1}{0}{0}{0} \tj{\ell}{\ell'}{1}{-m}{m'}{q}~.
\end{align}
In the expression for $R_q$, the first Wigner 3j-symbol can be simplified:
\begin{equation}
\tj{\ell}{\ell'}{1}{0}{0}{0} = 
\begin{cases}
(-1)^\ell \sqrt{\frac{\ell^2}{\ell(2\ell-1)(2\ell+1)}}~, & \text{if } \ell' = \ell -1~, \\
0~,  &\text{if } \ell' = \ell~, \\
(-1)^{\ell+1} \sqrt{\frac{(\ell+1)^2}{(\ell+1)(2\ell+1)(2\ell+3)}}~, & \text{if } \ell' = \ell +1~.
\end{cases}
\end{equation}
Therefore, we have the following 
\begin{align}
R_q(\ell m, \ell'm') &= 
\begin{cases}
(-1)^{\ell+m} \sqrt{(2\ell+1)(2\ell-1)}\sqrt{\frac{\ell^2}{\ell(2\ell-1)(2\ell+1)}} \tj{\ell}{\ell-1}{1}{-m}{m'}{q}~, & \text{if } \ell' =\ell -1~, \\
0~,  &\text{if } \ell' = \ell~, \\
(-1)^{\ell+m+1}\sqrt{(2\ell+1)(2\ell+3)} \sqrt{\frac{(\ell+1)^2}{(l+1)(2\ell+1)(2\ell+3)}}\tj{\ell}{\ell+1}{1}{-m}{m'}{q}~, & \text{if } \ell' = \ell +1~.
\end{cases} 
\\
&=
\begin{cases}
(-1)^{\ell+m} \sqrt{l} \tj{\ell}{\ell-1}{1}{-m}{m'}{q}~, & \text{if } \ell' = \ell -1~, \\
0~,  &\text{if } \ell' = \ell~, \\
(-1)^{\ell+m+1}\sqrt{\ell+1}\tj{\ell}{\ell+1}{1}{-m}{m'}{q}~, & \text{if } \ell' = \ell +1~.
\end{cases}
\end{align}
We then have:
\begin{itemize}
\item When $q = 1$,
\begin{align}
R_q(lm, \ell'm') &= 
\begin{cases}
\sqrt{\frac{(\ell+m-1)(\ell+m)} {2(2\ell-1)(2\ell+1)}} \delta_{m',m-1}, & \text{if } \ell' = \ell -1~, \\
0~,  &\text{if } \ell' = \ell~, \\
-\sqrt{\frac{(\ell-m+1)(\ell-m+2)} {2(2\ell+1)(2\ell+3)}}\delta_{m',m-1}, & \text{if } \ell' = \ell +1~.
\end{cases} 
\end{align}
\item When $q = 0$,
\begin{align}
R_q(lm, \ell'm') &= 
\begin{cases}
\sqrt{\frac{(\ell+m)(\ell-m)} {(2\ell-1)(2\ell+1)}}\delta_{m',m}, & \text{if } \ell' = \ell -1~, \\
0~,  &\text{if } \ell' = l~, \\
\sqrt{\frac{(\ell+m+1)(\ell-m+1)} {(2\ell+1)(2\ell+3)}}\delta_{m',m}, & \text{if } \ell' = \ell +1~.
\end{cases} 
\end{align}
\item When $q = -1$,
\begin{align}
R_q(lm, \ell'm') &= 
\begin{cases}
\sqrt{\frac{(\ell-m-1)(\ell-m)} {2(2\ell-1)(2\ell+1)}}\delta_{m',m+1}, & \text{if } \ell' = \ell -1~, \\
0~,  &\text{if } \ell' = \ell~, \\
-\sqrt{\frac{(\ell+m+1)(\ell+m+2)} {2(2\ell+1)(2\ell+3)}}\delta_{m',m+1}, & \text{if } \ell' = \ell +1~.
\end{cases} 
\end{align}
\end{itemize}
Then, because we take the real part when we calculate the force, the force matrix is 
$$
\mathbb P_q^{d} = \frac{\mathbb P_q^{d'} + (\mathbb P_q^{d'})^\dagger}{2},
$$
Note that we have the same block for $e-e$ and $h-h$ polarization. So for each pair of $\ell$ and $\ell'$, we need to have 2 copies of the matrix. 
\begin{figure} [htp]
    \centering
    \includegraphics[width=3.5in]{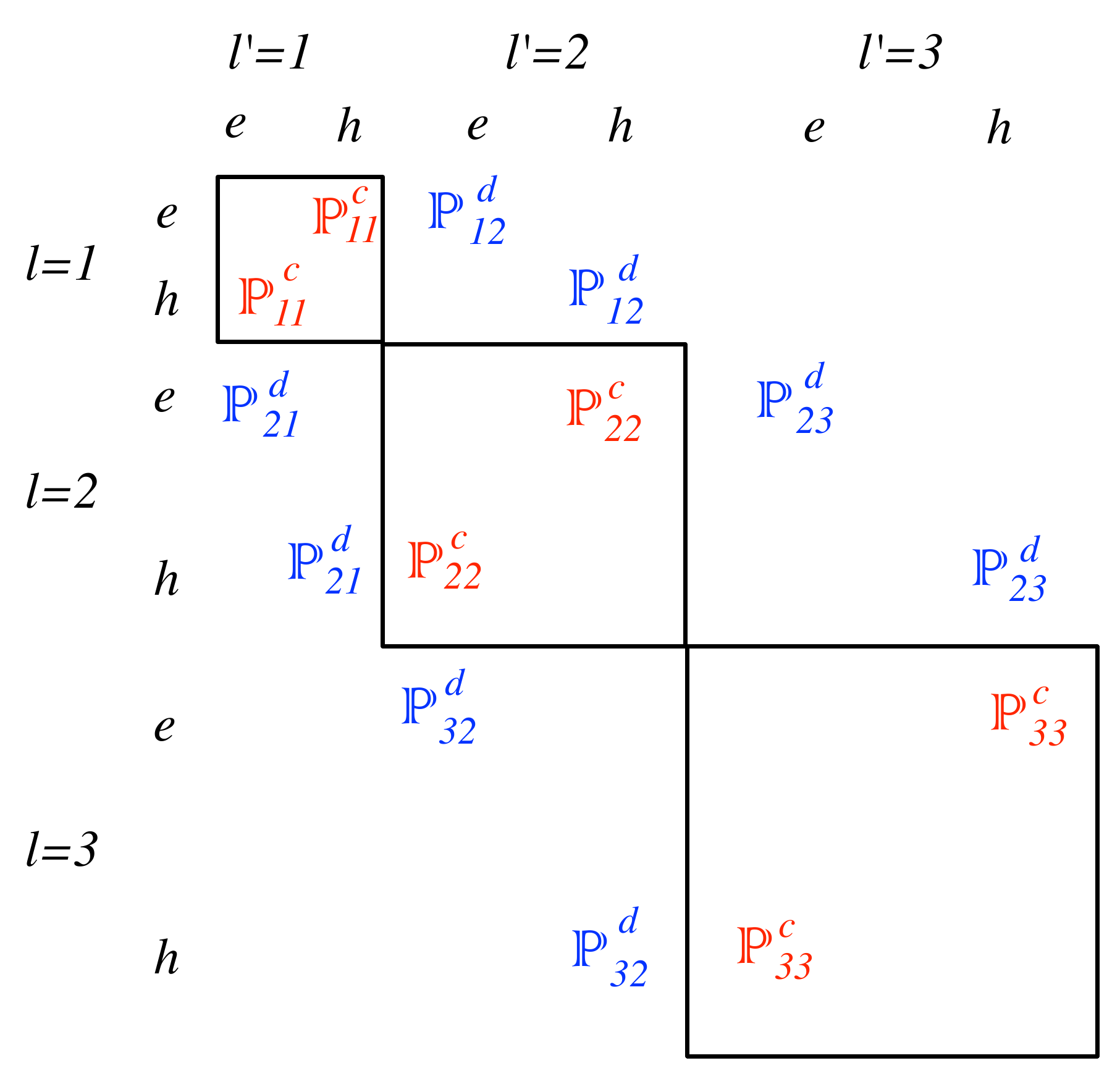}
    \caption{The structure of the force matrix $\mathbb P_z$}
    \label{fig:Fz}
\end{figure}

Now, let us focus on the $\mathbb P_q^c$. From \cite[Eqs.~(7.19,7.20)]{Farsund1996}, we have
\begin{equation}
\mathbb P_q^{c'} (\ell mm') = \frac{1}{\ell(\ell+1)} L_q(\ell mm')
\end{equation}
Note that this term is not along the diagonal since it is the $e$-$h$ interaction. $\mathbb P_q^{c'}$ has the same form for both $e-h$ and $h-e$ blocks. Again, there is a real operator for the force calculation, and therefore
\begin{equation}
\label{eq:Fc}
\mathbb P_q^{c} = \frac{\mathbb P_q^{c'} + (\mathbb P_q^{c'})^\dagger}{2}.
\end{equation}

Finally, $\mathbb P_q = \mathbb P_q^{d} + \mathbb P_q^{c}$. Using the same $q \rightarrow i$ conversion as for the torque case,
\begin{align}
\mathbb P_z &= \mathbb P_0 \\
\mathbb P_x &= \frac{\mathbb P_{+1} - \mathbb P_{-1}}{\sqrt{2}} \\
\mathbb P_y &= -i\frac{\mathbb P_{+1} + \mathbb P_{-1}}{\sqrt{2}}
\end{align}
$\mathbb P_z$, for example, has the structure shown in \figref{Fz}.  

\section{Bounds on eigenvalues of $\Pbb_i$ and $\Jbb_i$ in the VSW basis}
As we saw in \secref{torque_VSW}, the eigenvalues of the $\Jbb_i$, for any $i$, are simply the diagonal entries of $\Jbb_z$:
\begin{align}
    -\ell, -\ell+1, \ldots, \ell-1, \ell, \nonumber
\end{align}
and thus the maximum eigenvalue is
\begin{empheq}[box=\widefbox]{align}
    \lambda_{\rm max}(\Jbb_i) = \ell_{\rm max}.
\end{empheq}
For the force matrices $\Pbb_i$, the off-diagonal components make it impossible (as far as we can tell) to solve for the eigenvalues analytically. The Gershgorin circle theorem~\cite{Horn2013} can be used to get within about a factor of 1.5 of the largest eigenvalue, but it turns out that a simple physical argument yields a tighter bound.

\begin{figure} [htp]
    \centering
    \includegraphics[width=3.5in]{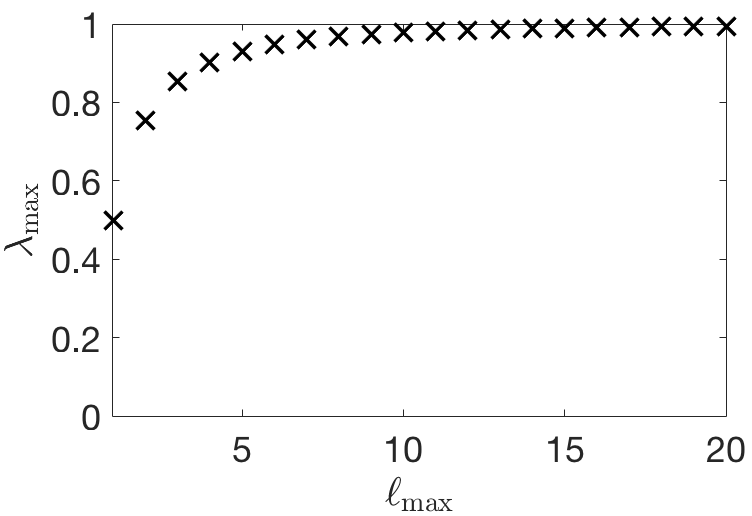}
    \caption{Largest eigenvalue of $\Pbb_i$ as a function of $\ell_{\rm max}$, converging to the bound of 1.}
    \label{fig:lambda_bound}
\end{figure}

Consider some set of incoming waves given by a set of coefficients $\cin$. The momentum per time carried by those waves is given by $\frac{1}{c} \cin^\dagger \Pbb_i \cin$. The maximum momentum that could be carried by those waves is given by the number of photons per unit time multiplied by $\hbar k$, i.e. the total momentum is less than or equal to the sum of $\hbar k = \hbar \omega / c$ for each photon. The number of photons per unit time is given by $\cin^\dagger \cin / \hbar \omega$, since $\cin^\dagger \cin$ is the incoming power. Following these arguments mathematically, we can write:

\begin{align*}
    \frac{1}{c} \cin^\dagger \Pbb_i \cin &\leq \hbar k \frac{dN}{dt} \\
    &= \hbar k \frac{\cin^\dagger \cin}{\hbar \omega} \\
    &= \frac{1}{c} \cin^\dagger \cin.
\end{align*}
We can rewrite the final expression without the speed of light,
\begin{align}
    \cin^\dagger \Pbb_i \cin \leq \cin^\dagger \cin,
\end{align}
which applies for \emph{any} $\cin$ vector, implying that
\begin{empheq}[box=\widefbox]{align}
    \lambda_{\rm max}(\Pbb_i) \leq 1.
\end{empheq}
Fig.~\ref{fig:lambda_bound} shows that the largest eigenvalue of $\Pbb_i$ in a VSW basis converges to the bound as $\ell_{\rm max} \rightarrow \infty$. Note that the eigenvalue bound itself does not rely on any property of VSWs and must be true for any basis.

\section{Plane-wave power and momentum in the VSW basis}
In this section we derive the quantities $\cin^\dagger \cin$, $\cin^\dagger \Pbb_i \cin$, and $\cin^\dagger \Jbb_i \cin$ in the case that $\cin$ represents the VSW coefficients for a plane wave propagating in the $z$ direction. We can start from the plane-wave expansions in \citeasnoun{Bohren1983} and convert to our VSW basis to find the incoming-wave coefficients. Plane waves have nonzero coefficients only for $m = \pm 1$; taking $m=1$ below, the coefficients for linear polarization are
\begin{subequations}
\begin{align}
    \vect{c}_{e\ell m}^{\rm Lin} &=  \frac{1}{2}\sqrt{\pi (2\ell-1)} i^{\ell-1} \\
    \vect{c}_{e\ell -m}^{\rm Lin}  &=  -\frac{1}{2}\sqrt{\pi (2\ell-1)} i^{\ell-1} \\
    \vect{c}_{h\ell m}^{\rm Lin}  &=  \frac{1}{2}\sqrt{\pi (2\ell+1)} i^{\ell-1} \\
    \vect{c}_{h \ell -m}^{\rm Lin}  &=  \frac{1}{2}\sqrt{\pi (2\ell+1)} i^{\ell-1},
\end{align}
\end{subequations}
for right circular polarization they are
\begin{subequations}
\begin{align}
    \vect{c}_{e\ell m}^{\rm RCP} &=  \frac{1}{2}\sqrt{2\pi (2\ell-1)} i^{\ell-1} \label{eq:RCP1}\\
    \vect{c}_{h\ell m}^{\rm RCP}  &=  \frac{1}{2}\sqrt{2\pi (2\ell+1)} i^{\ell-1}  \label{eq:RCP2},
\end{align}
\end{subequations}
and for left circular polarization they are
\begin{subequations}
\begin{align}
    \vect{c}_{e\ell -m}^{\rm LCP} &=  -\frac{1}{2}\sqrt{2\pi (2\ell-1)} i^{\ell-1} \\
    \vect{c}_{h\ell -m}^{\rm LCP}  &=  \frac{1}{2}\sqrt{2\pi (2\ell+1)} i^{\ell-1}.
\end{align}
\end{subequations}
The value of $\cin^\dagger \cin$ is the same for any polarization (since the power is not affected by polarization). It is simplest to compute the power for circular polarization, in which case it is the sum $\frac{1}{2} \sum_{\ell=1}^{\lmax} 2\pi \left(2\ell + 1\right)$. At this point we rescale our coefficients by the value $\frac{|\Ev_0|}{k \sqrt{2 Z_0}}$, where $\Ev_0$ is the plane-wave amplitude, $k$ the wavenumber, and $Z_0$ the impedance of free space (to ultimately yield a power-normalized $\cin^\dagger \cin$). Then,
\begin{empheq}[box=\widefbox]{align}
    \cin^\dagger \cin &= \frac{\pi}{k^2} \frac{|\Ev_0|^2}{2 Z_0} \left(\lmax^2 + 2\lmax\right).
\end{empheq}
All of the following quantities will ultimately be written in terms of $\cin^\dagger \cin$, so we drop the scale factor $|\Ev_0| / k\sqrt{2 Z_0}$ hereafter.

Now we consider the momentum flowing in direction $i$. The momentum per time is given by $(\hat{\vect{z}} \cdot \hat{\vect{i}}) \cin^\dagger \Pbb_z \cin/c = \beta_i \cin^\dagger \Pbb_z \cin/c$, since there is no $x$- or $y$-directed momentum. (It can be verified that $\cin^\dagger \Pbb_x \cin = \cin^\dagger \Pbb_y \cin = 0$.) From \secref{force_VSW}, we know that $\mathbb P_z = \mathbb P_z^d + \mathbb P_z^c$. We saw that
$$
\mathbb P_z^c(\ell mm') = \frac{1}{\ell(\ell+1)} L_0(\ell mm')~,
$$ 
which means that
\begin{align}
    \cin^\dagger \mathbb P_z^c \cin &= \sum_{\ell = 1} ^{\lmax} \frac{1}{\ell(\ell+1)} \cc{c_\ell} c_\ell \nonumber \\
                              &= \pi \sum_{\ell = 1} ^{\lmax } \frac{2\ell +1}{\ell(\ell+1)},
\end{align}
where we used the fact that $\cin$ has nonzero coefficients only for $m = 1$, for which $L_0(\ell m m') = \delta_{mm'}$. From \eqref{Fd}, the $\Pbb_z^d$ contribution is
\begin{align}
    \cin^\dagger \mathbb P_z^d \cin &= \sum_{\ell = 1} ^{\lmax-1} \frac{1}{2} \left[ (\ell^2 + \ell -1) + \left( \ell+1 \right)^2 + \ell + 1 -1 \right] \nonumber \\
&\cdot \sqrt{\frac{(\ell+2)\ell}{(2\ell+1)(2\ell+3)}}\cdot \sqrt{\frac{1}{\ell(\ell+1)}} \cdot \sqrt{\frac{1}{(\ell+1)(\ell+2)}} \nonumber \\
&\cdot \sqrt{2\pi(2\ell +1 )} \cdot \sqrt {2\pi (2\ell+3)} \nonumber \\
    &= \pi \sum_{\ell=1}^{\lmax-1} \frac{2\ell (\ell+2)}{(\ell +1 )}
\end{align}
The first line of the above equation is the summation of the $\{\ell,(\ell+1)\}$ and $\{(\ell+1),\ell\}$ interaction. For both interactions, $R_q(\ell m, \ell 'm')$ is the same. We vary $\ell$ from $1$ to $\lmax-1$ since the interaction only comes into play when $\lmax \geq 2$. The first term of the second line includes $R_q(\ell m, \ell'm')$. The third line incorporates the values of $\cin$ for channels $\ell$ and $\ell+1$. So the sum of the contributions from $\Pbb_z^c$ and $\Pbb_z^d$ is
\begin{align}
\cin^\dagger \mathbb P_z \cin &= \pi \sum_{\ell = 1} ^{\lmax-1}\frac{2\ell+1}{\ell(\ell+1)} + \pi \frac{2\lmax+1} {\lmax(\lmax+1)} + \pi \sum_{\ell=1}^{\lmax-1} \frac{2\ell(\ell+2)}{(\ell+1)} \nonumber \\
&=\pi \sum_{\ell = 1} ^{\lmax-1} \frac{2\ell(\ell+1)^2}{\ell(\ell+1)} + \pi \sum_{\ell = 1}^{\lmax -1} \frac{1}{\ell (\ell+1)}+ \pi \frac{2\lmax+1}{\lmax(\lmax+1)}  \nonumber \\
&= \pi \sum_{\ell = 1}^{\lmax-1} 2(\ell+1) + \pi \cdot \frac{\lmax-1}{\lmax} + \pi \frac{2\lmax+1}{\lmax(\lmax+1)} \nonumber \\
&= \pi \left( \lmax^2 + \lmax - 2 +\frac{\lmax-1}{\lmax} +  \frac{2\lmax+1}{\lmax(\lmax+1)} \right) \nonumber \\
&= \pi (\lmax^2 + 2\lmax)\frac{\lmax}{\lmax+1} \nonumber \\
&= \frac{\lmax}{\lmax+1} \cin^\dagger \cin.
\end{align}
And thus the momentum flow per time in direction $i$, denoted $\mathcal{P}_{\textrm{in},i}$ in the main text, is
\begin{empheq}[box=\widefbox]{align}
    \mathcal{P}_{\textrm{in},i} = \frac{\beta_i}{c} \frac{\lmax}{\lmax + 1} \cin^\dagger \cin.
\end{empheq}

Finally, for the angular momentum, we separately consider the RCP and LCP waves, and at the end show that the total angular momentum is proportional to the degree of right circular polarization. Again, one can show for any $\cin$ that $\cin^\dagger \Jbb_x \cin = \cin^\dagger \Jbb_y \cin = 0$, such that the angular momentum in direction $i$ is determined by the $z$-directed fraction,
\begin{align}
    \cin^\dagger \Jbb_i \cin = \beta_i \cin^\dagger \Jbb_z \cin.
\end{align}
For an RCP plane wave, the coefficients of $\cin$ are nonzero only for $m=1$, for which the diagonal entries of $\Jbb_z$ are $1$, such that $\cin^\dagger \Jbb_z \cin = \cin^\dagger \cin$. Conversely, for an LCP plane wave the coefficients of $\cin$ are nonzero only for $m=-1$, for which the diagonal entries of $\Jbb_z$ are $-1$, such that $\cin^\dagger \Jbb_z \cin = -\cin^\dagger \cin$, the negative of the RCP case. Thus is we define $\gamma_i$ as the degree of right circular polarization of any incoming wave, the angular momentum per unit time is
\begin{empheq}[box=\widefbox]{align}
    \mathcal{J}_{\textrm{in},i} = \frac{\beta_i \gamma_i}{\omega} \cin^\dagger \cin.
\end{empheq}

\section{Force bound when $\lmax = 1$}
In the main text, we derived force and torque bounds in a VSW basis for plane-wave incidence for any $\lmax$, using the eigenvalue bound $\lambda_{\rm max}(\Pbb_i) = 1$. Here, we consider the case $\lmax = 1$. In this case, analysis of the matrices in \secref{force_VSW} shows that $\lambda_{\rm max}(\Pbb_z) = 1/2$. Carrying this factor of 1/2 through the bound derivation, one finds that the force in the $i$ direction normalized by the incident-wave intensity is bounded above by
\begin{empheq}[box=\widefbox]{align}
    \frac{F_i}{I_{\rm inc}} &\leq \frac{3\lambda^2}{4\pi c},
\end{empheq}
about a factor of 1/3 tighter than the bound in the main text, for this special case.

\section{Helix: structural details}
The line running along the center of a helix wrapping around the $z$ axis has a simple parametrization:
\begin{align}
    \vect{r}(t) = (R\cos(t), R\sin(t), ht),
\end{align}
where $R$ controls the radius of that center line as it wraps, and $h$ scales the rate at which the height along $z$ changes. The parameter $t$ controls how many rotations of the helix occur, e.g. $[0, 4\pi]$ means two circles. For a three-dimensional helical structure, we need two unit vectors at each point along the center line, to create the circular surface slice of the helix. Starting with the tangent vector (by differentiation),
\begin{align}
    \vect{t}(t)  = (-R\sin(t), R\cos(t), h),
\end{align}
one can get the local normal vector as
\begin{align}
    \vect{n}(t) = (-\cos(t), -\sin(t), 0).
\end{align}
The second local basis vector is the ``binormal,''
\begin{align}
    \vect{b}(t) = \vect{t} \times \vect{n} =  \frac{1}{\sqrt{R^2 + h^2}} (h \sin(t), -h\cos(t), R). 
\end{align}
To create the 3D helix, we thus use a vector $\vect{S}$ that is the sum of $\vect{r}(t)$ with two new parameters and the two basis vectors. We use a parameter $u$ which ranges from 0 to $2\pi$, to create the circular surfaces around the helical line, and a second parameter $a$ that represents the radius of the circle that wraps around the center line (not the radius of the circle formed by the center line itself, which is $R$). 
\begin{align}
    \vect{S}(u,t) = \vect{r}(t) + a \vect{n}(t) \cos(u) + a \vect{b}(t) \sin(u). 
\end{align}
For the structure simulated in the main text, we used the values $R = 0.9$, $a = 0.45$, and $h = 0.3$.

\section{Cross-section bounds rederived}
In this final section we derive VSW bounds on scattering, absorption, and extinction cross-sections for plane-wave illumination, and verify that the resulting bounds agree with previous results from the literature~\cite{Hamam2007,Pozar2009,Ruan2011,Liberal2014,Hugonin2015}. To examine scattered power and extinction, we will need to connect the incoming-field/outgoing-field separation to the common incident-field/scattered-field separation. For VSWs, it is generally true for any incident field that half of the field must be incoming and the other half must be outgoing (to have a continuous field at the origin, where incoming/outgoing fields have singularities)~\cite{Doicu2006}. The scattered field must be purely outgoing. Thus, the relationship between the in/out coefficients $\cin$ and $\cout$, and the inc/scat coefficients $\cinc$ and $\cscat$ is given by \eqreftwo{in}{out}.
 
We start with absorption, the bound for which is particularly simple. Absorption is given by $\cin^\dagger \cin - \cout^\dagger \cout$, and thus maximum absorption satisfies
\begin{equation}
    \begin{aligned}
        & \underset{\cout}{\text{maximize}} & & \cin^\dagger \cin - \cout^\dagger \cout \\
        & \text{subject to}       & & \cout^\dagger \cout \leq \cin^\dagger \cin. 
    \end{aligned}
\end{equation}
Maximum absorption occurs when $\cout^\dagger \cout = 0$ (all power is incoming and absorbed), such that 
\begin{align}
    \Pabs^{\rm (max)} = \cin^\dagger \cin = \frac{\pi |\Ev_0|^2}{2 k^2 Z_0} \left(\lmax^2 + 2\lmax\right)
\end{align}
The absorption cross-section is the absorbed power divided by the incident intensity, $I_{\rm inc} = |\Ev_0|^2 / 2 Z_0$. Then the maximum absorption cross-section is
\begin{empheq}[box=\widefbox]{align}
    \sigma_{\rm abs}^{\rm (max)} = \frac{\pi}{k^2} \left(\lmax^2 + 2\lmax\right) = \frac{\lambda^2}{4\pi} \left(\lmax^2 + 2\lmax\right).
\end{empheq}

Scattered power is the outgoing power in the scattered fields, and hence is given by $\cscat^\dagger \cscat$. By \eqreftwo{in}{out}, $\cscat = \cout - \cin$, such that maximum scattered power is the solution to the optimization problem
\begin{equation}
    \begin{aligned}
        & \underset{\cout}{\text{maximize}} & & \left(\cout - \cin\right)^\dagger \left(\cout - \cin\right) \\
        & \text{subject to}       & & \cout^\dagger \cout \leq \cin^\dagger \cin. 
    \end{aligned}
\end{equation}
Lagrangian multipliers confirm the intuition that the optimal $\cout$ is the negative of $\cin$: $\cout = -\cin$. Then the scattered power will be $4\cin^\dagger \cin$, i.e. 4 times the maximum absorbed power, and the maximum scattering cross-section is
\begin{empheq}[box=\widefbox]{align}
    \sigma_{\rm scat}^{\rm (max)} = \frac{4\pi}{k^2} \left(\lmax^2 + 2\lmax\right) = \frac{\lambda^2}{\pi} \left(\lmax^2 + 2\lmax\right).
\end{empheq}

Extinction is the sum of the absorbed and scattered powers, and thus equals $2\Re \cin^\dagger \left(\cin - \cout\right)$ (which equals the more intuitive expression $\Re \cinc^\dagger \cscat$). Then the maximum extinction satisfies
\begin{equation}
    \begin{aligned}
        & \underset{\cout}{\text{maximize}} & & 2 \Re \cin^\dagger \left(\cin - \cout\right) \\
        & \text{subject to}       & & \cout^\dagger \cout \leq \cin^\dagger \cin. 
    \end{aligned}
\end{equation}
The maximum is achieved at the maximum-scattering condition, $\cout = -\cin$, meaning the extinction cross-section has the same upper bound as the scattering cross-section:
\begin{empheq}[box=\widefbox]{align}
    \sigma_{\rm ext}^{\rm (max)} = \frac{4\pi}{k^2} \left(\lmax^2 + 2\lmax\right) = \frac{\lambda^2}{\pi} \left(\lmax^2 + 2\lmax\right).
\end{empheq}

\bibliography{/Users/odm5/Library/texmf/bibtex/bib/library,/Users/odm5/Library/texmf/bibtex/bib/journalshort,/Users/odm5/Library/texmf/bibtex/bib/nonotes,/Users/odm5/Library/texmf/bibtex/bib/My_Pubs_abbr}